\documentclass{article} 

\PassOptionsToPackage{table}{xcolor}

\usepackage{iclr2027_conference,times}

\setcitestyle{numbers,square,comma}

\makeatletter
\def\addcontentsline#1#2#3{%
  \addtocontents{#1}{\protect\contentsline{#2}{#3}{\thepage}\protected@file@percent}}
\makeatother

\usepackage[T1]{fontenc}    
\usepackage{hyperref}       
\usepackage{url}            
\usepackage{booktabs}       
\usepackage{amsfonts}       
\usepackage{nicefrac}       
\usepackage{microtype}      
\usepackage{multirow}

\usepackage{graphicx}
\usepackage{xcolor}
\usepackage{amsmath}
\usepackage{amsthm}
\newtheorem{proposition}{Proposition}
\theoremstyle{definition}
\newtheorem{assumption}{Assumption}
\theoremstyle{plain}
\usepackage{wrapfig}
\usepackage{subcaption} 
\usepackage{algorithm}
\usepackage{algpseudocode}
\usepackage{enumitem}

\makeatletter
\newcommand{\appendixtableofcontents}{%
  \section*{Contents}
    \begingroup
    \small
    \setlength{\parskip}{2ex plus 0.2ex minus 0.2ex}%
    \setlength{\itemsep}{0.4ex}%
    \@starttoc{app}%
  \endgroup
}

\newcommand{\appsection}[1]{%
  \section{#1}%
  \addcontentsline{app}{section}{\protect\numberline{\thesection}#1}%
}

\newcommand{\appsubsection}[1]{%
  \subsection{#1}%
  \addcontentsline{app}{subsection}{\protect\numberline{\thesubsection}#1}%
}
\makeatother

\renewcommand{\thefootnote}{\fnsymbol{footnote}}

\title{\fontsize{16}{19}\selectfont Evolve as a Team: Collaborative Self-Evolution\\ for LLM-based Multi-Agent Systems}

\author{%
\bfseries
Zhezheng Hao\textsuperscript{1\,*},
Tianfu Wang\textsuperscript{2\,*},
Huanshuo Dong\textsuperscript{3},
Ziyan Liu\textsuperscript{3},
Hong Wang\textsuperscript{3},\\
\bfseries
Xiankun Lin\textsuperscript{3},
Qiang Lin\textsuperscript{3},
Can Wang\textsuperscript{1},
Hande Dong\textsuperscript{3\,\textdagger},
Jiawei Chen\textsuperscript{1\,\textdagger}
\\[1.0ex]
\mdseries
\textsuperscript{1}Zhejiang University
\,\,
\textsuperscript{2}The Hong Kong University of Science and Technology
\\
\textsuperscript{3}Workbuddy Team, Tencent
}

\iclrfinalcopy

\begin{document}

\maketitle


\begingroup
\renewcommand{\thefootnote}{\fnsymbol{footnote}}
\footnotetext[1]{Equal contribution.}
\footnotetext[2]{Corresponding author.}
\endgroup

\begin{abstract}
LLM-based multi-agent systems (MAS) have emerged as an effective paradigm for complex and long-horizon tasks.
However, in real-world tasks, MAS often exhibit various failures during execution and such failures are difficult to eliminate during design.
This motivates experience-driven MAS evolution, where a system improves based on its own execution experience.
Yet such evolution is challenging because MAS experience is prolonged and intricate, interleaving multiple agents’ execution chains and communication messages, which makes it difficult to identify what should be improved.
To address this challenge, we propose \textbf{Meta-Team}, an experience-driven MAS evolution framework based on collaborative self-evolution.
Meta-Team preserves the execution context of each agent and coordinates post-task communication, enabling agents to exchange distributed evidence for evolution.
Building on this design, Meta-Team conducts multi-scale self-evolution, transforming execution experience into reusable improvements to agent behaviors, inter-agent coordination, and team-level organization.
Across six long-horizon agent benchmarks, Meta-Team consistently outperforms single-agent systems, hand-crafted MAS, and prior MAS evolution methods; further analyses demonstrate that Meta-Team enables more reliable and scalable MAS self-evolution.\footnotetext[3]{Code is available at \url{https://github.com/zz-haooo/Meta-Team}.}
\end{abstract}

\section{Introduction}

Recent advancements in Large Language Models (LLMs)~\citep{gpt54, gemini3, anthropic2025claude46, liu2025deepseek} have driven rapid progress in LLM agents~\citep{wang2024openhands, openmanus2025, claude_code_overview_2026}. 
However, as LLM agents are increasingly applied to complex, long-horizon real-world tasks, a fundamental bottleneck of single-agent systems has emerged: they struggle to handle long-context information within a limited context window~\citep{hsieh2024ruler,liu2024lost} and face cognitive burden during long-horizon execution~\citep{zoureducing,shang2025united,an2025does}.

To address this bottleneck, LLM-based Multi-Agent Systems (MAS) have become an effective paradigm for complex and long-horizon tasks~\citep{guo2024large,li2024survey,kim2025towards}.
MAS leverage the ``wisdom of the crowd'' and follow a divide-and-conquer principle --- they decompose complex tasks into manageable subtasks, assign them to specialized agents, and coordinate the agents to complete the task collaboratively~\citep{zhang2024chain,zhao2024longagent,xu2025does}.
This paradigm effectively reduces per-agent context overload and makes complex tasks more tractable.
MAS have developed from early workflow-based systems~\citep{hong2023metagpt, qian2024chatdev, li2023camel} to recent industry-scale agent teams~\citep{team2026kimi, claude_code_agent_team}, demonstrating their growing utility in solving real-world tasks.

Despite this progress, building effective MAS still requires substantial human effort~\citep{anthropic2025research, zhou2025multi}.
Developers often need to manually design core MAS components, including agent prompts, interaction protocols, workflow structures, and coordination mechanisms, making such systems costly to build and difficult to adapt to diverse tasks.
Moreover, practical deployments show that MAS often exhibit failure modes such as inter-agent miscommunication, role drift, and cascading errors~\citep{phillips2026multiagent,tran2026single,yan2025dontbuild,papadopoulos2026mind}.
Since these failure modes arise from actual execution, they cannot be fully eliminated through design-time specifications alone.

These limitations have motivated recent studies on automated MAS generation and evolution. Early efforts primarily follow two lines:
(1) Performance-driven strategies, which leverage a meta-level optimizer to generate, search, or refine MAS configurations based on performance metrics (e.g., pass rates or final rewards)~\citep{huautomated, shangagentsquare, zhuge2024gptswarm, zhang25aflow, zhang2025multi};
(2) Memorization-driven strategies, which directly store MAS trajectories or interaction fragments for subsequent retrieval~\citep{li2025cross,lin2025se, yang2025agentnetdecentralizedevolutionarycoordination}.
However, both paradigms lack fine-grained attribution over MAS experience: they provide limited insight into why and how a system succeeds or fails, leaving evolution to rely on exhaustive trial-and-error or superficial trajectory reuse rather than targeted improvement.

More recently, to bridge this attribution gap, studies have employed a single centralized LLM analyzer to inspect full MAS trajectories, pinpoint failures, and propose revisions for MAS evolution~\citep{vuautomated26,hu2026evolutionary,liu2026mas}.
While promising, this single-analyzer paradigm remains highly unreliable on long trajectories: existing failure-attribution benchmarks show that centralized analysis performs poorly in pinpointing the decisive failure step, and the performance declines as trajectories grow longer~\citep{zhang2025agent,zhang2025agentracer,banerjee2025did}.
We argue that this limitation stems from a fundamental architectural mismatch between MAS execution and MAS evolution ---  while MAS mitigate context overload during execution by distributing task information across agents, existing evolution methods regress by flattening the entire team's trajectory into a single context for one analyzer.
Consequently, they reintroduce the single-context bottleneck that MAS were originally designed to overcome. This bottleneck is even more pronounced during evolution, given that the full team trajectory is particularly prolonged and intricate~\citep{li2026towards,chen2026seeing}.

In this work, we advocate a simple principle for MAS evolution: \textbf{a MAS should not only execute as a team, but also evolve as a team}.
Following this principle, we propose \textbf{Meta-Team}, a MAS evolution framework based on collaborative self-evolution.
Embracing the philosophy of \emph{team reflexivity} in organizational psychology~\citep{west1996reflexivity,schippers2015team}, Meta-Team preserves local execution contexts within the agents that produced them and coordinates inter-agent communication so that agents can exchange information required for evolution.
This collaborative organization matches the distributed structure of MAS experience, avoiding the context and reasoning burden of flattening the full MAS trajectory into a single analyzer.
Specifically, Meta-Team conducts self-evolution at three levels:
(i) \emph{Agent-level evolution}, where each agent reflects on its own execution behavior and updates its individual agent scaffold;
(ii) \emph{Interaction-level evolution}, where agents revisit their collaboration history to refine communication and teammate profiles;
and (iii) \emph{Team-level evolution}, where the team revises its organization and shared coordination rules via collective discussion.
Instead of evaluating on saturated benchmarks, we conduct experiments on six challenging long-horizon agent benchmarks, where Meta-Team consistently improves over existing agents, hand-crafted MAS, and prior MAS evolution methods.
For example, Meta-Team outperforms hand-crafted MAS by $6.6\%$ on average across the nine benchmark columns in Table~\ref{tab:main_results}.
These results demonstrate the effectiveness of collaborative self-evolution across diverse long-horizon tasks.


In summary, our main contributions are:
\begin{itemize}
    \item We identify an architectural mismatch in experience-driven MAS evolution: MAS execute through distributed contexts, whereas existing evolution methods centralize the full experience, making it difficult to attribute failures and identify what to improve.

\item We propose \textbf{Meta-Team}, a collaborative self-evolution framework that preserves agent-local contexts, coordinates post-task communication to exchange distributed evidence, and translates the distributed evidence into multi-scale evolution.

    \item We conduct extensive experiments and analyses. These results validate the effectiveness and scalability of Meta-Team in MAS self-evolution.
\end{itemize}

\section{Preliminaries}
\subsection{LLM-based MAS}
An LLM-based Multi-Agent System (MAS) consists of multiple LLM agents that collaborate to solve a task.
We define a MAS as $$
\mathcal{M}\equiv(\mathcal{A},\mathcal{O}),
$$
where \(\mathcal{A}=\{a_i\}_{i=1}^{N}\) is a set of agents and \(\mathcal{O}\) denotes the orchestration mechanism that specifies how the agents are organized and how they interact and coordinate during task execution.
Each agent \(a_i\) consists of a backbone LLM $\phi$ and an agent scaffold \(s_i\), including its role instruction, task strategy, tools, and memory mechanism~\citep{shangagentsquare,hao2026recreate}.
The orchestration mechanism $\mathcal{O}$ can take various forms in existing systems~\citep{hong2023metagpt,wu2024autogen,li2024survey}.
For clarity, we decompose the orchestration mechanism as
$\mathcal{O}=(\mathcal{P},\mathcal{S})$,
where $\mathcal{P}$ denotes inter-agent protocols and $\mathcal{S}$ denotes the shared team scaffold.
The protocol $\mathcal{P}$ specifies when and how agents exchange information and hand off intermediate results.
For instance, workflow-based MAS often follow predefined information flows, whereas flexible agent-team systems allow agents to reason about when to contact teammates and what information to share.
The shared team scaffold $\mathcal{S}$ specifies team-level rules visible to all agents, such as the team roster, organization, and shared constitution.

\subsection{Experience-Driven Self-Evolution of MAS}
Given a task $x$, executing a MAS $\mathcal{M}$ yields an execution experience defined as
$$
e\equiv(x,\tau,r),
$$
where $\tau$ is the multi-agent trajectory and $r$ is the task outcome, such as the final deliverable and evaluator score.
Unlike a single-agent trajectory, $\tau$ is not a single execution chain, but an interleaving of multiple agents' local execution chains, communication messages, and intermediate artifacts.
We denote the local execution chain of agent $a_i$ as $\tau_i$, and write the complete trajectory as
$$
\tau=\operatorname{Interleave}(\tau_1,\ldots,\tau_N,\mathcal{B}),
$$
where $\mathcal{B}$ denotes cross-agent events such as messages and shared artifacts.
This formulation highlights the distributed structure of MAS experience: each agent holds a local execution context, while the complete team trajectory also depends on cross-agent interactions.

Experience-driven self-evolution aims to improve a MAS from its own execution experience across tasks~\citep{shinn2023reflexion,hao2026recreate,liu2026mas}.
Given a sequence of experiences $\mathcal{E}_{1:k}=\{e_1,\ldots,e_k\}$, the MAS is updated by
$$
\mathcal{M}_{k+1}=\Omega(\mathcal{M}_k,\mathcal{E}_{1:k}).
$$
Since task outcomes are usually sparse, self-evolution requires determining what should be improved from the trajectory and applying changes to the MAS.

\section{Method}
\subsection{Collaborative Scheme for MAS Self-Evolution}
Real-world tasks often call for long-context understanding and long-horizon interaction~\citep{qiu2025locobench,li2026agencybench}.
Solving them requires agents to perform hundreds of interaction steps and consume millions of context tokens, imposing a heavy burden of long-context understanding and long-horizon reasoning on a single agent.
MAS alleviate this dual burden by distributing task information and execution responsibility across multiple agents during execution, allowing each agent to focus on more manageable subtasks within shorter and more coherent contexts~\citep{zhang2024chain,zhao2024longagent,xu2025does,shang2025united}.

However, the distributed execution creates a new challenge for self-evolution: the resulting experience is longer and more complex than any single agent's local trajectory, making it difficult to identify what should be improved~\citep{kong2025aegis,zhang2025agentracer,chen2026seeing}.
Existing analyzer-based evolution methods follow a \emph{Global Scheme}: they flatten this joint experience and feed it to a single centralized LLM analyzer to attribute failures and propose revisions~\citep{hu2026evolutionary,liu2026mas}.
While providing a global view, the global scheme concentrates the context and reasoning burden back on a single analyzer, reintroducing the bottleneck that MAS were originally designed to alleviate.
A natural and naive counterpart is \emph{Local Scheme}: decomposing the analysis by agent, with each analyzer reflecting on the corresponding local trajectory to identify local failures and propose agent-specific revisions, in the spirit of single-agent reflection~\citep{shinn2023reflexion,agrawal2025gepa,hao2026recreate}.
Although the local scheme reduces the per-analyzer context length, it lacks a global view to understand how agents' local decisions, messages, and dependencies jointly shape the final outcome, making it difficult to identify what should truly be improved.

\begin{figure}[t]
    \centering
    \begin{subfigure}[t]{0.62\textwidth}
        \centering
        \includegraphics[width=\textwidth, height=5.2cm]{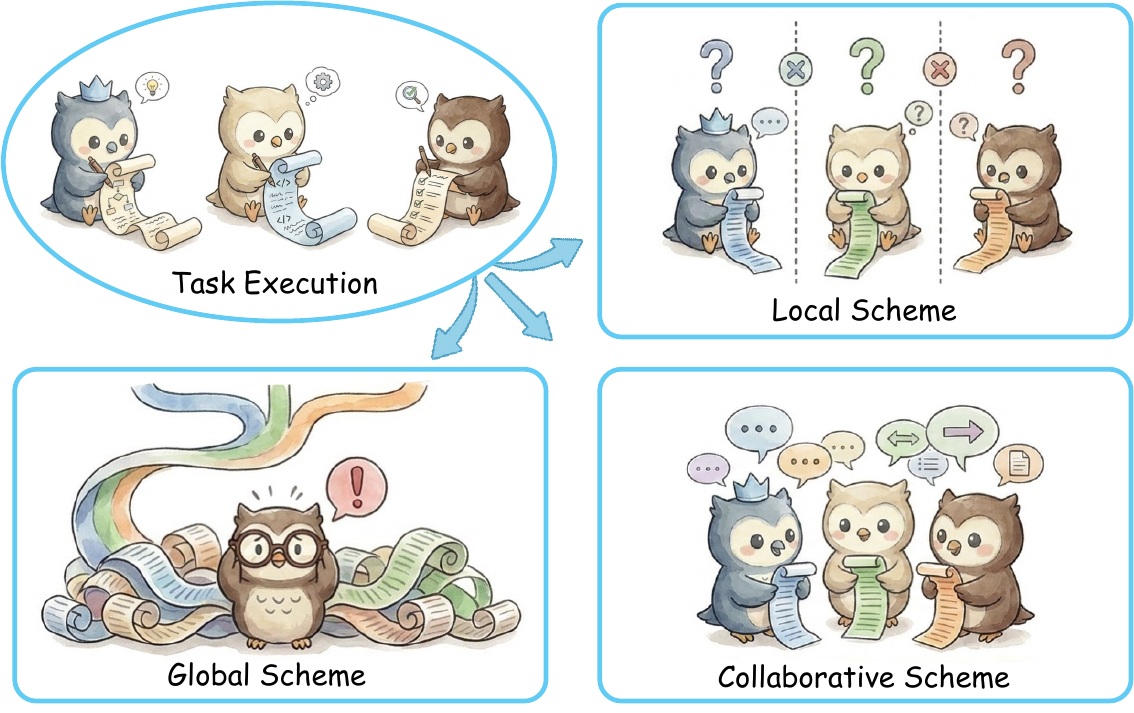}
        \caption{Three schemes for organizing MAS experience during evolution.}
        \label{fig:schematic}
    \end{subfigure}
    \hfill
    \begin{subfigure}[t]{0.34\textwidth}
        \centering
        \includegraphics[width=\textwidth]{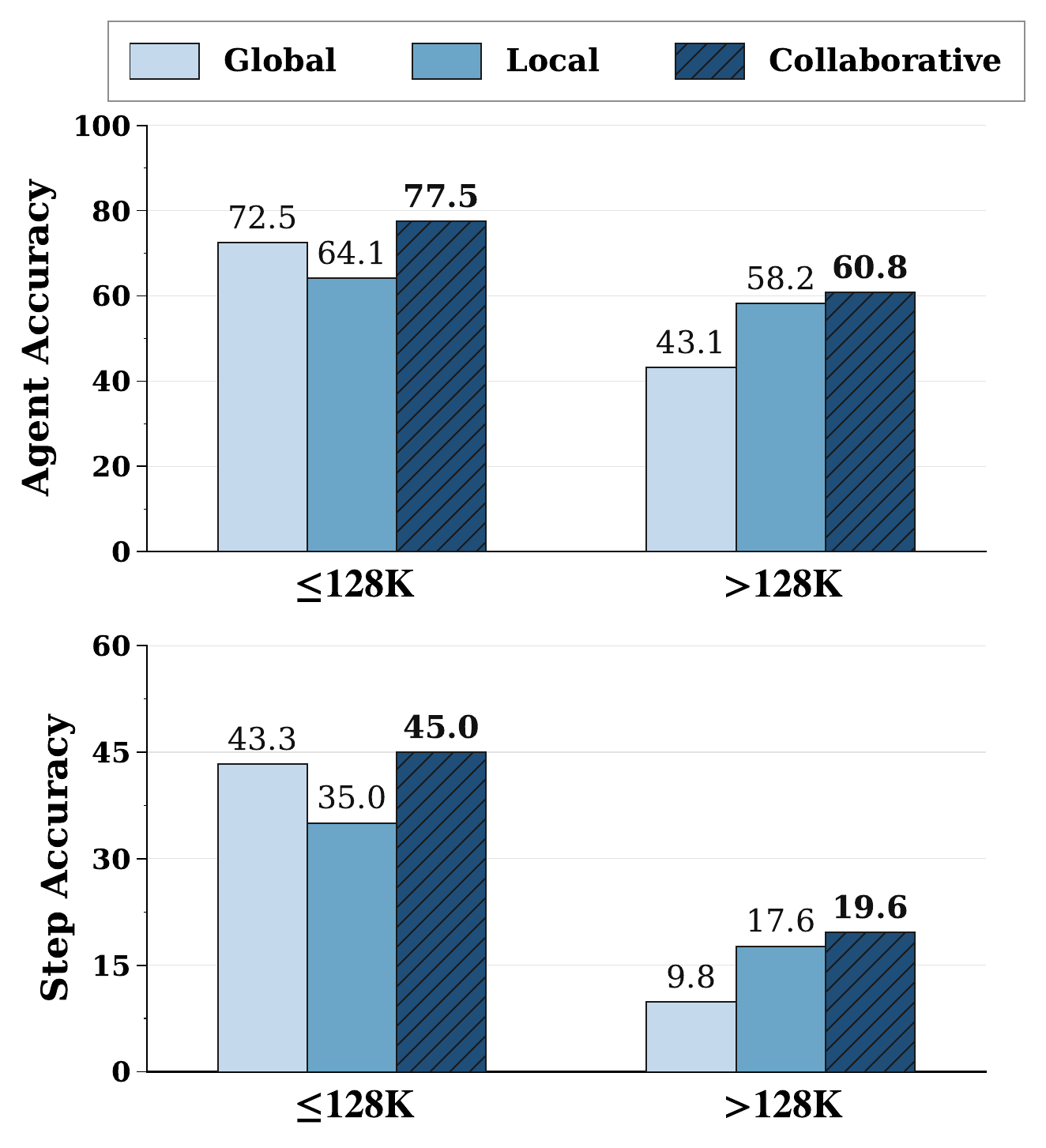}
        \caption{Failure-attribution Comparison.}
        \label{fig:meta-team-overview-right}
    \end{subfigure}
    \vspace{-0.4em}
    \caption{Three different schemes of experience-driven MAS self-evolution.}
    \label{fig:global-local-collaborative}
    \vspace{-1.5em}
\end{figure}

This reveals a global-local trade-off in organizing MAS experience for self-evolution.
The global scheme preserves cross-agent coverage yet imposes a heavy centralized burden; the local scheme preserves local fidelity yet lacks cross-agent awareness.
To address this trade-off, we propose a \emph{Collaborative Scheme} for MAS self-evolution: each agent carries its local execution context into the evolution stage, while agents actively communicate to exchange processed local findings and trace how their outputs affected downstream execution.
In this way, collaborative experience organization preserves the fidelity of agent-local contexts while recovering the cross-agent awareness needed to identify reliable targets for MAS evolution.
Figure~\ref{fig:schematic} illustrates how the global, local, and collaborative schemes organize MAS experience for self-evolution.

\paragraph{Empirical validation.}
To validate the global-local trade-off and the benefit of collaborative experience organization, we compare the three schemes in a failure-attribution setting.

\emph{Settings.}
We evaluate the global, local, and collaborative schemes on TraceElephant~\citep{chen2026seeing}, a failure-attribution benchmark containing $220$ real multi-agent failure traces from three public MAS frameworks.
Each trace is annotated with a ground-truth \((\texttt{mistake\_agent}, \texttt{mistake\_step})\) tuple, and the goal is to identify the failure-causing agent (Agent Accuracy) and the decisive failure step (Step Accuracy).
The three schemes differ only in how they organize the same MAS trajectory for attribution: global uses the full flattened trajectory, local uses isolated agent-wise trajectories, and collaborative uses agent-wise trajectories with cross-agent exchange.
All schemes use the same backbone LLM, Claude Sonnet 4.6, and we report avg@3 Agent Accuracy and Step Accuracy.
Implementation details are provided in Appendix~\ref{app:three-schemes-exp}.

\emph{Findings.}
The results are shown in Figure~\ref{fig:global-local-collaborative}, where the dataset is split by trajectory length into \(\leq 128\mathrm{K}\) and \(>128\mathrm{K}\) regimes to examine how context length affects different schemes.
Notably, all trajectories are within the \(1\mathrm{M}\)-token context limit of the backbone LLM, so the global scheme always has access to the full trajectory without truncation. On trajectories shorter than \(128\mathrm{K}\), the global scheme outperforms the local scheme, suggesting that full-trajectory access is useful when the context remains manageable. However, on trajectories longer than \(128\mathrm{K}\), the global scheme drops sharply and falls behind the local scheme on both metrics, indicating that long interleaved MAS traces impose a severe context and reasoning burden even without context overflow. By contrast, the collaborative scheme consistently achieves the best performance in both length regimes.
These results show that, compared with purely global or local analysis, the collaborative scheme better aligns with the distributed structure of MAS experience and provides more reliable guidance for self-evolution.

Beyond empirical analysis, we give a theoretical analysis of the global, local, and collaborative schemes in Appendix~\ref{app:theory}, which shows that the global--local ordering flips with trajectory length while the collaborative scheme dominates both regardless of length.

\paragraph{Discussion.} Beyond its benefit for identifying failure causes through context decomposition, collaborative self-evolution reframes agents from passive targets of diagnosis into active contributors to system evolution.
This collaborative philosophy brings two further advantages for MAS evolution.

\emph{Endogenous Feedback.} Existing evolution methods often rely on final task outcomes or verifier scores as feedback~\citep{hao2026recreate,yang2025agentnetdecentralizedevolutionarycoordination,liu2026mas}, which are too sparse to specify how agents should improve their behavior.
Collaborative self-evolution enriches this feedback through post-task communication: agents explain how others' outputs affected their decisions and receive feedback on how their own outputs affected downstream execution.
In this way, the team's interactions become endogenous feedback signals for agent evolution.

\emph{Bottom-Up Self-Evolution.}
Collaborative self-evolution also enables agents to surface system-level improvements.
Since agents observe different parts of the execution, they can identify team-level issues such as missing roles, redundant responsibilities, unclear handoffs, or ineffective coordination rules.
Following the principle of team reflexivity~\citep{west1996reflexivity,schippers2015team}, the team can aggregate these observations through collective discussion and revise its composition and organization.
In this way, distributed execution experience becomes the basis for bottom-up self-evolution of the whole MAS.

\subsection{Meta-Team: Multi-Scale Collaborative Self-Evolution}
\label{sec:method:metateam}
\begin{figure}[t]
    \centering
    \includegraphics[width=\textwidth]{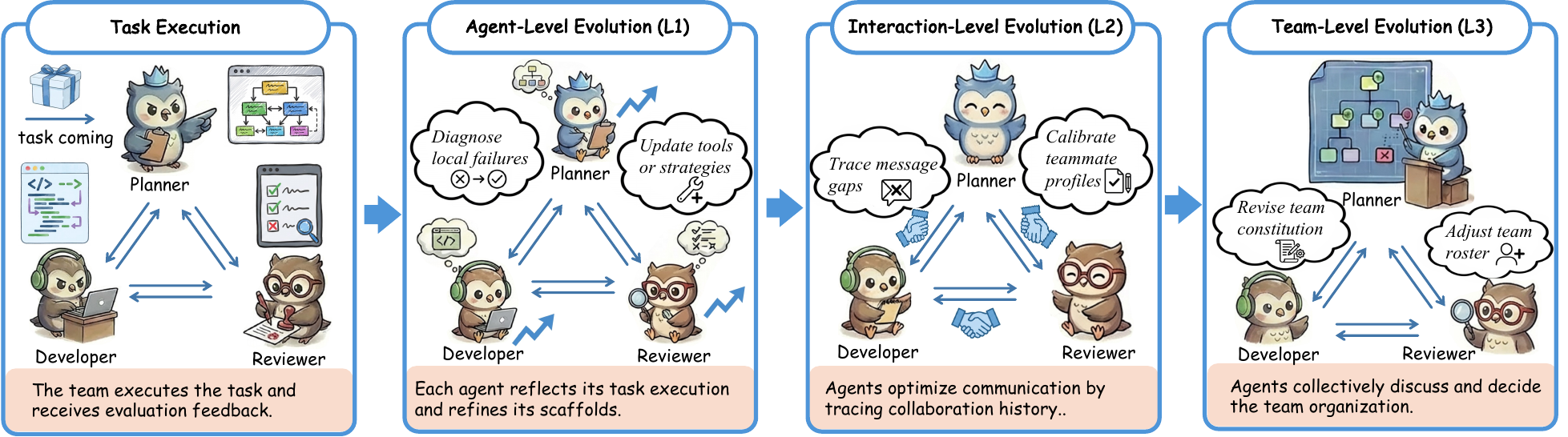}
    \vspace{-0.8em}
\caption{Overview of Meta-Team.}
    \label{fig:meta-team-overview}
    \vspace{-1.5em}
\end{figure}

The previous subsection establishes how MAS experience should be organized for self-evolution: agent-local contexts are preserved and connected through post-task communication.
Meta-Team instantiates this principle by aligning evolution with the compositional structure of MAS.
After each task, agents receive the final deliverable and evaluation result, collaboratively revisit the execution process, and convert distributed evidence into updates at three complementary scopes: agent-level behavior, interaction-level collaboration, and team-level coordination.
These scopes are complementary: a single failure may simultaneously reveal an agent-specific limitation, a cross-agent dependency issue, and a team-level coordination gap~\citep{lin2025agentask,in2026rethinking}.
The overall process is illustrated in Figure~\ref{fig:meta-team-overview}.

\paragraph{Agent-Level Evolution (L1).}
At the agent level, the evolution targets are individual agent scaffolds $s_i$.
Although the update is applied to one agent $a_i$, it is derived from collaborative self-evolution rather than from that agent's local context alone.
The agent reviews evidence from its own execution chain and actively solicits cross-agent evidence to verify how its assumptions, intermediate outputs, and decisions affected downstream execution.
This enables agent-specific improvements with both local fidelity and cross-agent awareness, such as what the agent should observe, verify, report, or avoid in future tasks.

\paragraph{Interaction-Level Evolution (L2).}
At the interaction level, the evolution target is the collaboration mechanism $\mathcal{P}$ between agents.
Through communication, agents revisit their previous collaboration to examine how information flowed and how agents relied on one another.
Rather than optimizing a binary communication graph~\citep{yang2025agentnetdecentralizedevolutionarycoordination,zhou2025multi}, Meta-Team updates agents' teammate profiles, which capture how agents understand, query, and rely on one another.
These profiles help agents calibrate their understanding of each teammate's strengths and limitations, enabling more effective collaboration in subsequent tasks.

\paragraph{Team-Level Evolution (L3).}
At the team level, the evolution target is the shared team scaffold $\mathcal{S}$, including the team roster, the organizational structure among agents, and the shared team constitution, i.e., the global prompt that defines common objectives, collaboration principles, and decision rules for all members.
Through collective discussion and decision-making, Meta-Team examines whether the team has the appropriate composition, organization, and shared rules for the task.
It then decides whether to introduce new roles, remove redundant ones, reorganize coordination, or revise the shared constitution.
In this way, Team-Level Evolution improves the team's overall operating mechanism rather than only patching individual agents or pairwise interactions.

After the three-scale evolution, Meta-Team commits validated updates to the reusable team scaffold, including agent patches, teammate profiles, collaboration notes, and revisions to the shared constitution.
An implementation pipeline is provided in Appendix~\ref{app:pipeline}, including the prompts, tools, and per-level procedure of the collaboration mechanism and a worked example (Appendix~\ref{app:pipeline:mechanism} and~\ref{app:pipeline:example}).

\section{Experiments and Results}
In this section, we conduct our experiments from the following perspectives.
First, we compare Meta-Team against nine strong baselines on six agent benchmarks to examine its overall effectiveness (\S\ref{sec:Main Results}).
Second, we conduct ablation studies to isolate the contribution of Meta-Team's core designs, including collaborative self-evolution and multi-scale evolution (\S\ref{sec:exp:ablation}).
Third, we study scalability and generalization by evaluating whether Meta-Team remains effective across increasing context lengths and generalizes beyond the evolution distribution (\S\ref{sec:Scalability and Generalization of Meta-Team}).
Finally, we analyze its efficiency under constrained budgets, testing whether Meta-Team achieves robust performance with limited time and cost (\S\ref{sec:Evolution under Constrained-Budget}).
Further experiments on cross-model generalization and the stability of the evolution process are reported in Appendix~\ref{app:additional_results}.

\subsection{Experimental Setup}
\label{sec:exp:setup}
\paragraph{Benchmarks.}
Although prior MAS and MAS-evolution methods have been evaluated on various benchmarks, many of these benchmarks have become saturated under rapidly advancing LLM capabilities, detailed in Appendix~\ref{app:old_benchmarks}.
Instead, we evaluate Meta-Team on six challenging benchmarks: \textbf{SWE-bench Pro}~\citep{deng2025swe} and \textbf{BeyondSWE}~\citep{chen2026beyondswe} for realistic software-engineering tasks, \textbf{LOCA-Bench}~\citep{zeng2026loca} for long-context productivity assistants with MCP-style tools, \textbf{GAIA}~\citep{mialon2023gaia} for multi-step open-web reasoning, \textbf{LoCoBench}~\citep{qiu2025locobench} for long-context repository-level coding, and \textbf{ResearchRubrics}~\citep{sharma2025researchrubrics} for open-ended research evaluation.
Appendix~\ref{app:benchmarks} provides detailed benchmark descriptions, experimental settings, and initial configurations for each benchmark.

\paragraph{Baselines.}
We compare Meta-Team against three families of baselines; implementation details are in Appendix~\ref{app:baselines}.
\emph{(i) Hand-crafted agents and multi-agent systems:}
\textbf{SA} (Single-Agent)~\citep{yao2023react}, a ReAct executor under the same tools and budget as Meta-Team; \textbf{MAS}, our hand-designed scaffold identical to Meta-Team's initial configuration but without evolution; \textbf{AggAgent}~\citep{lee2026agentic}, which aggregates $K$ independent trajectories into a final answer; \textbf{OWL}~\citep{huowl}, a fixed Planner--Coordinator--Worker hierarchy; and \textbf{AOrchestra}~\citep{ruan2026aorchestra}, which synthesizes sub-agents at runtime via a central orchestrator.
\emph{(ii) Performance-driven search:} \textbf{AgentSquare}~\citep{shangagentsquare}, which searches a modular Planning/Reasoning/Tool-Use/Memory design space under a scalar reward.
\emph{(iii) Experience-based evolution} (our closest comparison points): \textbf{ReCreate}~\citep{hao2026recreate}, which creates domain-specialized agents from past histories; \textbf{AgentNet}~\citep{yang2025agentnetdecentralizedevolutionarycoordination}, a decentralized DAG of agents with retrieval-augmented, self-evolving edges; and \textbf{MASFly}~\citep{liu2026mas}, which adapts the MAS at test time using a distilled SOP and experience pool revised by a single analyzer.


\paragraph{Evolution and Evaluation Setup.}
Following~\citep{wu2024prompt,hao2026recreate}, we hold out around 20 instances per benchmark or subset as the evolution set.
All evolution methods (AgentSquare, ReCreate, AgentNet,  MASFly, Meta-Team) consume only this set during evolution and are evaluated on the disjoint held-out split with all learned artifacts frozen.
Unless stated otherwise, all experiments use Claude Sonnet~4.6~\citep{sonnet} as the base LLM (\texttt{temperature}~$=$~0.2, \texttt{max\_tokens}~$=$~32768).
Full event traces (LLM calls, tool calls, inter-agent messages) are recorded for experience-driven evolution.
To balance evaluation cost and variance, all results on all selected benchmarks are reported as avg@3 following~\citep{hu2026evolutionary, liu2026mas}.

\begin{table*}[t]
\centering
\definecolor{ourrow}{RGB}{235,243,252}
\caption{Main results across diverse agent benchmarks. The best results are in \textbf{bold}.}
\label{tab:main_results}
\vspace{-0.5em}

{
\fontsize{8.8pt}{9.0pt}\selectfont
\setlength{\tabcolsep}{4pt}
\renewcommand{\arraystretch}{1.13}
\newcommand{\best}[1]{\mathbf{#1}}
\newcommand{\na}{\text{--}}
\begin{tabular}{@{}lcccccccccc@{}}
\toprule
\multirow{2}{*}{\textbf{Method}}
& \multicolumn{2}{c}{\textbf{SWE-Pro}}
& \multicolumn{2}{c}{\textbf{BeyondSWE}}
& \multicolumn{1}{c}{\textbf{LOCA}}
& \multicolumn{1}{c}{\textbf{GAIA}}
& \multicolumn{2}{c}{\textbf{LoCoBench}}
& \multicolumn{1}{c}{\textbf{ResRub.}}
& \multicolumn{1}{c}{\multirow{2}{*}{\textbf{Avg.}}} \\
\cmidrule(lr){2-3}\cmidrule(lr){4-5}\cmidrule(lr){6-6}\cmidrule(lr){7-7}\cmidrule(lr){8-9}\cmidrule(lr){10-10}
& \textbf{Ansible}
& \textbf{Qute.}
& \textbf{DepMig.}
& \textbf{CrossR.}
& \textbf{Val}
& \textbf{Val}
& \textbf{Feat.}
& \textbf{Refact.}
& \textbf{All}
& \\
\midrule
SA
& $44.7$ & $62.7$ & $43.5$ & $41.7$ & $78.3$ & $70.0$ & $64.5$ & $62.0$ & $43.9$ & $56.8$ \\
MAS
& $40.8$ & $61.0$ & $37.6$ & $40.4$ & $82.1$ & $74.7$ & $57.6$ & $60.9$ & $49.5$ & $56.1$ \\
AggAgent
& $49.6$ & $57.7$ & $42.8$ & $40.6$ & $80.4$ & $70.3$ & $60.2$ & $54.7$ & $47.2$ & $55.9$ \\
OWL
& $39.5$ & $60.1$ & $36.7$ & $36.9$ & $72.9$ & $64.0$ & $53.8$ & $56.2$ & $47.9$ & $52.0$ \\
AOrchestra
& $44.7$ & $54.0$ & $44.1$ & $42.0$ & $82.9$ & $73.3$ & $62.0$ & $65.5$ & $51.5$ & $57.8$ \\
AgentSquare
& $39.0$ & $56.3$ & $38.4$ & $37.4$ & $68.3$ & $65.0$ & $56.2$ & $55.9$ & $41.9$ & $50.9$ \\
ReCreate
& $42.5$ & $59.6$ & $41.6$ & $39.1$ & $74.6$ & $69.0$ & $58.9$ & $62.0$ & $45.9$ & $54.8$ \\
AgentNet
& $40.8$ & $58.7$ & $44.3$ & $39.6$ & $69.6$ & $67.3$ & $57.0$ & $56.9$ & $40.3$ & $52.7$ \\
MASFly
& $43.4$ & $62.9$ & $39.7$ & $39.4$ & $75.8$ & $71.0$ & $60.1$ & $64.5$ & $50.8$ & $56.4$ \\
\rowcolor{ourrow}
\textbf{Meta-Team}
& $\best{53.9}$ & $\best{66.2}$ & $\best{45.6}$ & $\best{43.3}$ & $\best{87.9}$ & $\best{77.3}$ & $\best{67.1}$ & $\best{67.0}$ & $\best{55.8}$ & $\best{62.7}$ \\
\bottomrule
\end{tabular}
}

\vspace{0.25em}
\begin{minipage}{0.98\textwidth}
\fontsize{7.2pt}{8.4pt}\selectfont
\textit{Abbreviations.} Ans.~=~ansible, Qute.~=~qutebrowser, DepMig.~=~DepMigrate, CrossR.~=~CrossRepo, Feat.~=~Feature Implementation, Refact.~=~Cross-File Refactoring, ResRub.~=~ResearchRubrics. \textbf{Avg.} is the unweighted arithmetic mean across the nine columns.
\end{minipage}
\vspace{-1em}
\end{table*}

\subsection{Main Results}\label{sec:Main Results}
Table~\ref{tab:main_results} reports the main results across six challenging agent benchmarks.
Meta-Team achieves the best performance on all evaluated benchmark columns.
This suggests that the benefit of Meta-Team is not tied to a single task format, but rather generalizes across heterogeneous agent settings.

We first compare Meta-Team with the single-agent SA baseline and the initial hand-designed MAS.
The initial MAS uses the same starting team scaffold as Meta-Team, but does not evolve from execution experience.
Notably, the initial MAS is not consistently better than the single-agent baseline: it underperforms single-agent on six out of nine benchmark columns, suggesting that a human-designed MAS may suffer from organizational or coordination issues.
In contrast, after collaborative self-evolution, Meta-Team outperforms both single agent and the MAS on every evaluated column.
The gains against MAS are particularly clear on long-horizon tasks, such as SWE-bench Pro Ansible with $+13.1$ points and LoCoBench Feat. with $+9.5$ points.
This shows that Meta-Team does not merely rely on having multiple agents; rather, it turns an initially imperfect team into an adaptive system whose organization and coordination improve from execution experience.

Meta-Team also outperforms prior automated and experience-driven evolution methods, which exploit task feedback or past trajectories through performance-driven search, memory retrieval, or centralized analyzer-based revision.
Meta-Team obtains consistent gains over these baselines, including an average improvement of $6.3$ points over MASFly, the strongest MAS evolution baseline.
This comparison shows that simply using experience is not sufficient; how the experience is organized is critical.
The consistent gains of Meta-Team indicate that preserving agent-local contexts and enabling post-task communication provides a more effective way to convert distributed execution experience into MAS improvements.
All results are reported as avg@3; we verify the reliability of these margins with paired significance tests (Appendix~\ref{app:statistical_reliability}) and report their variance across independent evolution runs (Appendix~\ref{app:additional:stability}).

\begin{table}[t]
\centering
\begin{minipage}[t]{0.48\linewidth}
\vspace{0pt}
\centering
\captionof{table}{Ablation on collaborative self-evolution.}
\label{tab:ablation_collab}
\vspace{-0.5em}
\small
\setlength{\tabcolsep}{0pt}
\begin{tabular*}{\linewidth}{@{\extracolsep{\fill}}lcc}
\toprule
\textbf{Experience schemes} & \textbf{Ansible} $\uparrow$ & \textbf{ResRub.} $\uparrow$ \\
\midrule
No evolution (MAS)              & $40.8$ & $49.5$ \\
Centralized experience          & $49.8$ & $52.9$ \\
Partitioned experience          & $44.5$ & $51.0$ \\
\textbf{Collaborative experience} & $\mathbf{53.9}$ & $\mathbf{55.8}$ \\
\bottomrule
\end{tabular*}
\end{minipage}
\hfill
\begin{minipage}[t]{0.48\linewidth}
\vspace{0pt}
\centering
\captionof{table}{Ablation on multi-scale evolution.}
\label{tab:ablation_scales}
\vspace{-0.5em}
\small
\setlength{\tabcolsep}{0pt}
\begin{tabular*}{\linewidth}{@{\extracolsep{\fill}}lcc}
\toprule
\textbf{Variant} & \textbf{Ansible} $\uparrow$ & \textbf{ResRub.} $\uparrow$ \\
\midrule
\textbf{Full Meta-Team}  & $\mathbf{53.9}$ & $\mathbf{55.8}$ \\
\,\,w/o L1 (agent)       & $48.5$ & $47.9$ \\
\,\,w/o L2 (interaction) & $50.7$ & $54.4$ \\
\,\,w/o L3 (team)        & $51.9$ & $52.1$ \\
\bottomrule
\end{tabular*}
\end{minipage}
\vspace{-1.8em}
\end{table}

\subsection{Ablation Study}
\label{sec:exp:ablation}

We conduct ablations on SWE-bench Pro Ansible and ResearchRubrics to examine two core designs of Meta-Team: collaborative self-evolution and multi-scale evolution.
All settings are kept identical to the main experiments, except for the experience scheme or the enabled evolution scales.

Table~\ref{tab:ablation_collab} compares different schemes of organizing execution experience for MAS evolution.
Centralized experience refers to flattening the whole-team trajectory into a single LLM analyzer call for evolution.
Partitioned experience refers to spawning one LLM analyzer for each agent on its local experience, without cross-agent information exchange.
Collaborative experience refers to Meta-Team, which preserves the agent-wise partition of execution contexts and coordinates agents to exchange local information based on their reasoning abilities.
Although centralized and partitioned experience improve over no evolution, they remain clearly below collaborative experience.
This result is consistent with the attribution analysis in Figure~\ref{fig:global-local-collaborative}: collaborative experience yields more reliable evolution targets by combining agent-local evidence with cross-agent discussion.
The resulting improvement suggests that Meta-Team's gains come not merely from using past trajectories, but from organizing them in a form better suited to MAS failure attribution and self-evolution.

Table~\ref{tab:ablation_scales} evaluates the three evolution scales.
Removing any scale hurts performance, indicating that the three evolution scales are complementary to the MAS design.
L1 is the most influential scale: removing agent-level evolution drops performance by $5.4$ points on Ansible and $7.9$ points on ResearchRubrics, suggesting that improving individual agent scaffolds is the most direct lever for MAS improvement.
L2 and L3 show benchmark-dependent effects: L2 contributes more on Ansible, where structured software tasks rely on precise coordination, while L3 contributes more on ResearchRubrics, where open-ended evaluation benefits from better team organization.
Together, these ablation studies demonstrate that Meta-Team's gains come from both collaborative experience analysis and multi-scale evolution.

\begin{figure}[t]
    \centering
    \begin{subfigure}[t]{0.49\textwidth}
        \centering
        \includegraphics[width=\textwidth]{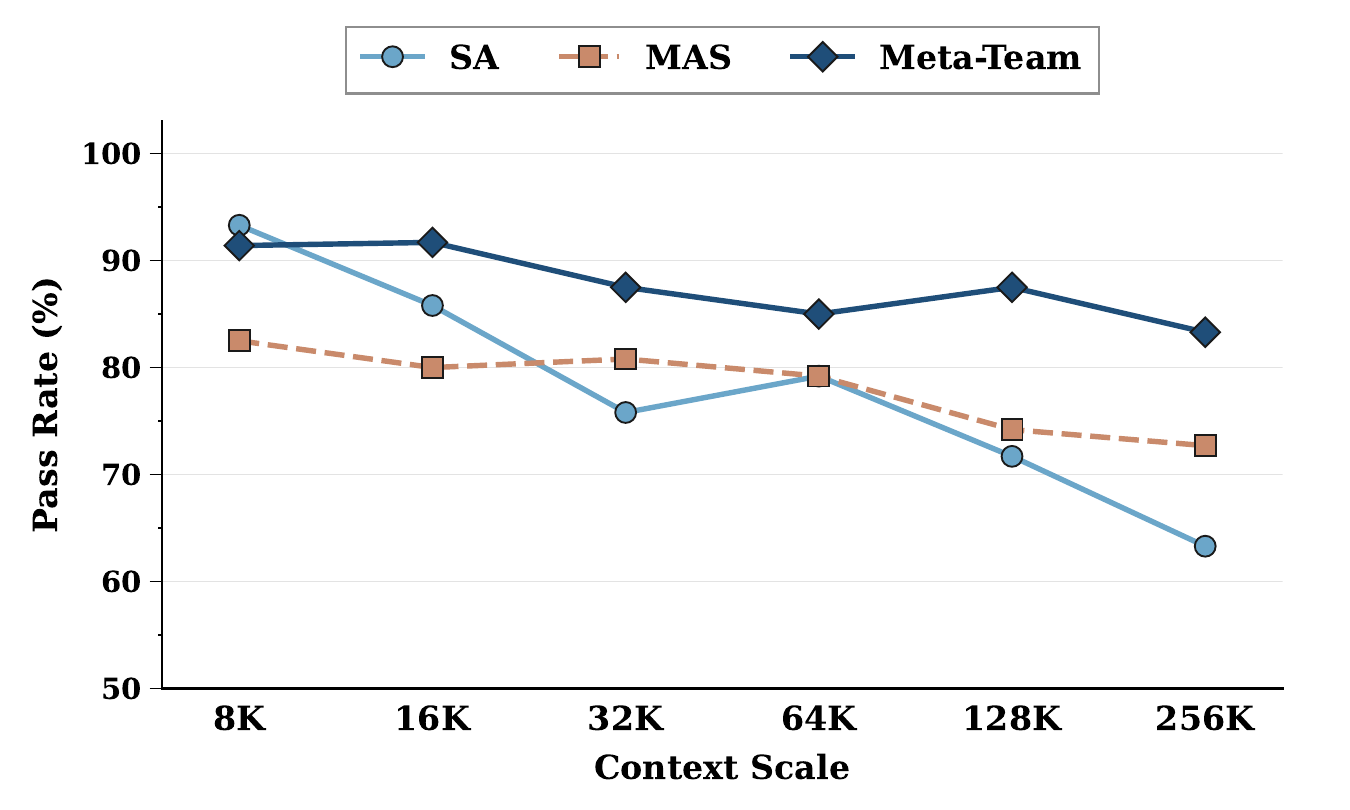}
        \caption{Experiments with different context scales.}
        \label{fig:locabench_scale}
    \end{subfigure}
    \hfill
    \begin{subfigure}[t]{0.49\textwidth}
        \centering
        \includegraphics[width=\textwidth]{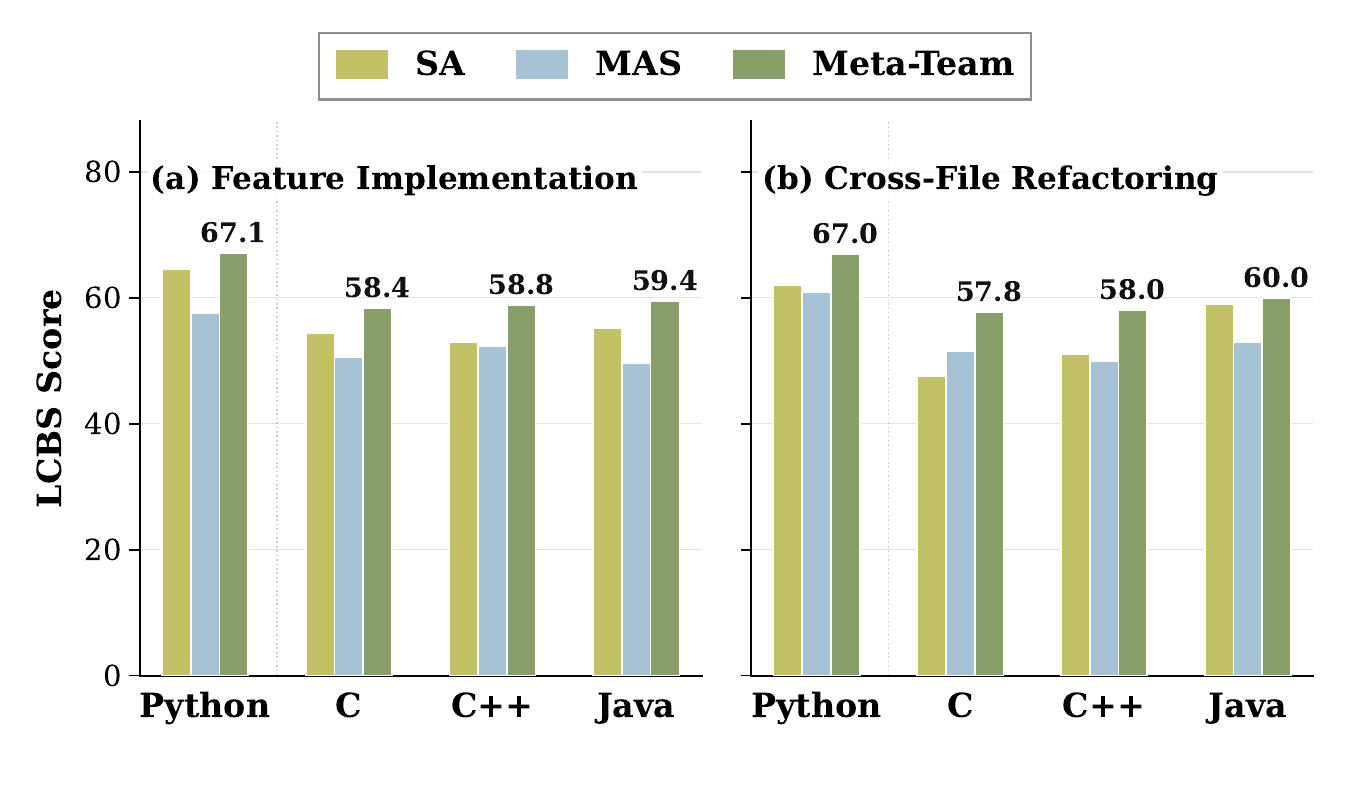}
        \caption{Experiments with various evaluation sets.}
        \label{fig:locobench_crosslang}
    \end{subfigure}
    \vspace{-0.4em}
    \caption{Experiments on scaling context and out-of-distribution evaluation sets.}
    \label{fig:locobench_combined}
    \vspace{-0.8em}
\end{figure}

\begin{figure}[t]
    \centering
    \includegraphics[width=\textwidth]{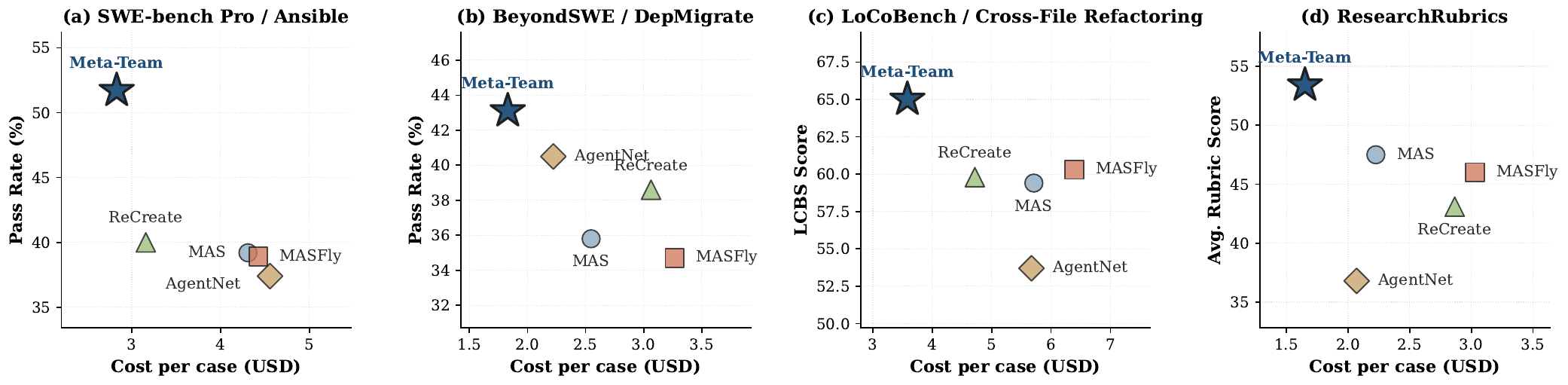}
    \vspace{-1em}
\caption{Experiments with constrained budget.}
    \label{fig:constrained_budget}
    \vspace{-1.5em}
\end{figure}

\subsection{Scalability and Generalization of Meta-Team}\label{sec:Scalability and Generalization of Meta-Team}
We further examine whether Meta-Team remains effective beyond the distribution used for evolution.
We study two axes: scalability across context length on LOCA-Bench and out-of-distribution generalization across programming languages on LoCoBench.
In LOCA-Bench, Meta-Team is evolved only on 96K-token cases with held-out seeds, and is evaluated on standard seeds from 8K to 256K.
In LoCoBench, Meta-Team is evolved on Python tasks and evaluated on C, C++, and Java tasks.

Figure~\ref{fig:locabench_scale} shows the scalability results on LOCA-Bench.
As the context length increases, the single-agent baseline degrades substantially, while the fixed MAS is more stable but remains below Meta-Team.
Meta-Team maintains the best performance across all context lengths except 8K context.
This suggests that collaborative self-evolution does not merely improve performance at the evolution context length, but learns more robust coordination patterns that continue to help under extreme context growth.
Interestingly, Meta-Team introduces two additional workers during evolution on 96K context, which explains why Meta-Team is especially effective in the longest-context setting (256K), where both the single-agent baseline and the fixed MAS face stronger context-management pressure.

Figure~\ref{fig:locobench_crosslang} reports the cross-language results on LoCoBench.
Although evolution is performed on Python tasks, Meta-Team consistently outperforms both the single-agent baseline and the fixed MAS on C, C++, and Java for both Feature Implementation and Cross-File Refactoring.
This indicates that the evolved improvements are not limited to language-specific shortcuts.
Instead, Meta-Team learns reusable collaboration behaviors, such as task decomposition, handoff, and verification, that transfer to unseen programming languages.

\subsection{Evolution under Constrained Budget}\label{sec:Evolution under Constrained-Budget}

In real deployment, MAS are running with time, cost, and token limits.
Therefore, we examine whether Meta-Team can evolve effective teams under a limited cost budget.
For all evolution methods, we constrain the evolution budget to $1/3$ of the main setting, freeze the resulting artifacts, and evaluate them under the same per-case deployment limits.
We compare MAS, ReCreate, AgentNet, MASFly, and Meta-Team on four benchmarks.
The evaluation results are shown in Figure~\ref{fig:constrained_budget}, where the x-axis reports the average inference cost per evaluation case.
The results show that Meta-Team achieves the best performance-cost trade-off across all four settings.
It obtains the highest performance with relatively low average evaluation cost, outperforming the other MAS self-evolution methods.
This advantage arises because Meta-Team can use timeout feedback on the evolution set to adjust the team system, allowing it to better adapt to the constrained-budget setting.
These results suggest that collaborative self-evolution helps Meta-Team learn more cost-efficient team behaviors under constrained evolution budgets.
A detailed accounting of the evolution cost, including a comparison with evolution baselines, is provided in Appendix~\ref{app:additional:cost}.

\vspace{-1em}

\section{Related Work}
\vspace{-0.5em}
\subsection{LLM-based Multi-Agent Systems}
MAS coordinate multiple language agents through role specialization, communication, and workflow-level organization. Early systems such as CAMEL~\citep{li2023camel}, ChatDev~\citep{qian2024chatdev}, MetaGPT~\citep{hong2023metagpt}, and AutoGen~\citep{wu2024autogen} demonstrate that specified roles, protocols, and workflows can improve collaborative problem solving across tasks such as software development, reasoning, and tool use.
More recent systems extend this paradigm to tool-rich and long-horizon settings~\citep{yang2024swe,wang2024openhands,openmanus2025}. For example, Magentic-One~\citep{fourney2024magentic} coordinates specialized agents through a central orchestrator, while OWL~\citep{huowl} adopts a workforce-style organization with planning, coordination, and execution roles. Recent agent-team systems, including agent swarm~\citep{team2026kimi}, agent teams~\citep{claude_code_agent_team}, and AOrchestra~\citep{ruan2026aorchestra}, further show the practical value of multi-agent coordination in real-world task automation. These works mainly study how agents should be organized during task execution, whereas our work studies how such organizations can be updated from execution experience.

\subsection{Automated MAS Generation and Evolution}
Although LLM-based MAS have shown promise, recent studies show that they remain brittle in realistic settings, with failures often stemming from poor specification, system-design flaws, inter-agent misalignment, and weak verification or termination mechanisms~\citep{cemrimulti,zhang2025agent}.
Other studies further show that small errors at inter-agent handoffs can propagate into system-level failures~\citep{lin2025agentask}, and that self-organizing agent teams may fail to effectively leverage their strongest members~\citep{pappu2026multi}.
These challenges motivate automated methods for designing, adapting, and evolving MAS.
Recent work has explored automated generation and optimization of agent systems. One line of work formulates MAS construction as performance-driven search over prompts, modules, workflows, communication graphs, or routing policies. Representative methods include ADAS~\citep{huautomated}, AgentSquare~\citep{shangagentsquare}, GPTSwarm~\citep{zhuge2024gptswarm}, AFlow~\citep{zhang25aflow}, MASS~\citep{zhang2025multi}.
Subsequent work trains workflow optimizers \citep{zhang2025multi,li2025agentswift,gao2025flowreasoner,wang2025scoreflow,yue2025masrouter,chen2025scoring,zhang2026flowsteer,kong2026workflow}. These methods optimize agent-system configurations using task-level feedback and reducing human design effort.
Inspired by experience learning in LLM and single-agent~\citep{shinn2023reflexion,madaan2023self,agrawal2025gepa,hao2026recreate}, another line of work uses execution experience to support MAS self-evolution.
Cross-task experiential learning methods store MAS trajectories or interaction fragments for later retrieval~\citep{li2025cross}, while AgentNet~\citep{yang2025agentnetdecentralizedevolutionarycoordination} evolves decentralized agent connectivity and expertise through accumulated experience.
ASpec~\citep{vuautomated26} cultivates persistent specialist teams for adaptive agent systems, and MASFly~\citep{liu2026mas} adapts MAS at test time by leveraging experience-guided revision.
Meta-Team follows this experience-driven direction, but organizes evolution according to the distributed structure of MAS itself, enabling agents to preserve local execution contexts while collaboratively evolving the system.

\section{Conclusion}
\label{sec:conclusion}
In this paper, we study how LLM-based multi-agent systems can improve from their own execution experience.
We advocate a simple principle for MAS evolution: a multi-agent system should not only execute as a team, but also evolve as a team. Building on this principle, we propose Meta-Team, which preserves agent-local contexts, enables post-task communication, and transforms distributed execution experience into targeted improvements for the team. We believe Meta-Team paves the way toward self-evolving agent teams that continuously improve their behaviors, communication, and coordination from their own distributed experience.





\bibliography{references}
\bibliographystyle{unsrtnat}


\clearpage

\appendix
\appendixtableofcontents

\appsection{Limitations}
This work has two main limitations.

First, Meta-Team focuses on scaffold- and organization-level evolution, including agent behaviors, collaboration patterns, and team-level coordination rules. It does not adapt the underlying execution infrastructure, such as tool APIs, environment harnesses, verifiers, or memory backends, which typically require task-specific engineering.

Second, Meta-Team does not update the parameters of the underlying LLMs. Its evolution relies on post-task reflection and explicit scaffold updates, making it interpretable and model-agnostic, but limiting the extent to which recurring team behaviors can be internalized into the base models. Combining collaborative self-evolution with fine-tuning or reinforcement learning is a promising future direction.

\appsection{Broader Impact}
Meta-Team introduces a collaborative self-evolution paradigm for LLM-based multi-agent systems by enabling agent teams to improve from their own distributed execution experience. Instead of relying on static, manually designed architectures or centralized post-hoc analyzers, Meta-Team allows agents to refine their roles, collaboration patterns, and team-level coordination through structured reflection. This approach can reduce the human effort required to design and maintain effective MAS, while improving their adaptability on long-horizon and tool-rich tasks. By making multi-agent systems more self-improving, inspectable, and reusable, Meta-Team has the potential to broaden access to intelligent automation across applications such as software engineering, research assistance, education, and enterprise workflows.

\appsection{Analysis of the Global-Local-Collaborative Schemes}\label{app:global-local-collaborative}

\appsubsection{Experimental Comparison}\label{app:three-schemes-exp}

To examine this design choice, we compare three attribution paradigms on TraceElephant, a benchmark of 220 real MAS failure traces from captain-agent~\citep{song2024adaptive}, magentic-one~\citep{fourney2024magentic}, and swe-agent~\citep{yang2024swe}.

\paragraph{Dataset.}
TraceElephant~\citep{chen2026seeing} provides $220$ failed multi-agent traces collected from three open-source MAS frameworks:
Captain-Agent~\citep{song2024adaptive} ($n{=}85$, short analytical tasks),
Magentic-One~\citep{fourney2024magentic} ($n{=}91$, Magentic-One-style orchestrated tasks) and SWE-Agent~\citep{yang2024swe} ($n{=}44$, SWE-Agent traces on \textsc{SWE-bench Verified}).
Each trace is a time-ordered list of step records (agent name, full input, full output), ranges from $2$ to $94$ steps (median $21$) involving $2$--$4$ agents, and is annotated with a ground-truth $(\texttt{mistake\_agent},\texttt{mistake\_step})$ tuple.
Our split yields $106$ traces in the $\le\!128$K split and $114$ in the $>\!128$K split.
Agent-Accuracy and Step-Accuracy are exact-match rates against the annotated tuple, averaged per split by simple arithmetic mean.

\paragraph{Settings.}
The three schemes are implemented as follows.

The global scheme issues a single LLM call on the flattened trace and reads the $(\texttt{agent},\texttt{step})$ pair directly from its JSON response.

The local scheme spawns one analyzer per agent, each seeing only its own sub-trace and submitting $(\texttt{i\_erred}\in\{0,1\},\,\texttt{my\_step},\,c)$, where $c\in[0,1]$.
The final $(\texttt{agent},\texttt{step})$ is the submission of ($\texttt{i\_erred}{=}1$) with the largest confidence $c$; if nobody self-accuses, we charge the first step of the least-confident denier, exposing the ``no coordination'' failure mode.

The collaborative scheme adds multiple rounds of communication: each analyzer first posts a short summary of its own findings, reads the summaries posted by the other analyzers, and then re-audits its sub-trace, submitting a tuple 
$$
(\texttt{i\_erred},\texttt{my\_step},c,r)
$$
,where $c\!\in\![0,1]$ is its self-confidence and $r\!=\!1$ if it disagrees with the global verdict and backs the disagreement with explicit counter-evidence from its own sub-trace.
Each analyzer then votes for the pair it accuses with weight $w=c\,(1+\alpha r)$, where $\alpha{=}1$ and the highest-scoring pair wins.

\appsubsection{Theoretical Analysis}
\label{app:theory}

We now analyze the global, local, and collaborative schemes formally. Using a
simple model of how an
analyzer recovers evidence from a trajectory, we show why the collaborative
scheme outperforms the global and local schemes across trajectory lengths, and
why the ranking between the latter two flips as the trajectory grows.

\paragraph{Setup.}
Consider a MAS with $N$ agents. Their steps are interleaved into one trajectory
of total length $L$, and agent $i$'s own part has length $\ell_i$, so
$L=\sum_{i=1}^{N}\ell_i$. To find the decisive cause of a failure, an analyzer
reads a trajectory (or part of it) and looks for the relevant clues. We describe
this with three quantities:
\begin{itemize}[leftmargin=1.5em,itemsep=1pt,topsep=2pt]
\item \emph{Local evidence} $S=\sum_{i=1}^{N}s_i$ (with $s_i \ge 0$): clues that
sit inside a single agent's own part of the trajectory.
\item \emph{Interaction evidence} $B \ge 0$: clues that only appear when one
agent hands off to another, i.e.\ they live between agents rather than inside
any one of them.
\item \emph{Recovery fraction} $\rho(\ell) \in (0,1]$: the share of the clues an
analyzer actually picks up when it reads a context of length $\ell$.
\end{itemize}
Getting the attribution right means recovering as many of the $S+B$ clues as
possible. We write $V$ for the \emph{effective evidence} a scheme recovers ---
the amount of $S+B$ its analyzers actually pick up, after dilution --- which is
the quantity the schemes will differ in. The analysis rests on the
following two assumptions.

\begin{assumption}[distributed evidence]
\label{ass:distributed}
Each agent's own part is much shorter than the whole trajectory,
$\ell_i \ll L$, and the interaction evidence $B$ comes from how different agents'
parts fit together, so it cannot be found by looking at any single agent's part
alone.
\end{assumption}
The short-part condition is just what dividing the work among agents gives us.
The second half says $B$ really is about the links between agents, which is why
a scheme that only looks inside each agent's own part can never see it.

\begin{assumption}[context dilution]
\label{ass:dilution}
The recovery fraction $\rho$ does not increase as the context gets longer: the
more an analyzer has to read, the smaller the share of clues it actually picks
up.
\end{assumption}
This is the well-documented effect that models use long contexts less
effectively. It does not need any cutting-off of the input: even if the whole
trajectory fits in the context window, a longer $\ell$ still lowers
$\rho(\ell)$.

\paragraph{Evidence recovered by each scheme.}
We compare the schemes by their effective evidence $V$: since accuracy is
non-decreasing in $V$ ($\mathrm{Acc}=g(V)$), the scheme that recovers more
evidence is at least as accurate. In the collaborative scheme an analyzer
additionally reads a teammate's brief summary of length $\delta$, so its context
$\ell^\star=\max_i \ell_i + \delta$ is still short ($\ell^\star \ll L$). The
schemes then obtain
\begin{align}
V_{\mathrm{global}} &= \rho(L)\,(S+B), \label{eq:v-global}\\
V_{\mathrm{local}}  &= \sum_{i=1}^{N}\rho(\ell_i)\,s_i, \label{eq:v-local}\\
V_{\mathrm{collab}} &= \sum_{i=1}^{N}\rho(\ell_i)\,s_i + \rho(\ell^\star)\,B. \label{eq:v-collab}
\end{align}
Intuitively, the global scheme sees $B$ but dilutes everything by $\rho(L)$; the
local scheme keeps its short-context fidelity but misses $B$ entirely; the
collaborative scheme keeps the local terms and adds $B$ back at only the mild
dilution $\rho(\ell^\star)$.

\begin{proposition}
\label{prop:flip}
There is a critical length $L_c$ such that $V_{\mathrm{global}} \ge
V_{\mathrm{local}}$ for short trajectories ($L \le L_c$) and $V_{\mathrm{global}}
< V_{\mathrm{local}}$ for long ones ($L > L_c$).
\end{proposition}
\begin{proof}
Subtracting \eqref{eq:v-local} from \eqref{eq:v-global},
\begin{equation}
f(L)\;:=\;V_{\mathrm{global}}-V_{\mathrm{local}}
=\rho(L)\,B-\sum_{i=1}^{N}\big[\rho(\ell_i)-\rho(L)\big]s_i. \label{eq:flip}
\end{equation}
When the trajectory is short, every $\ell_i$ is short too, so each bracket
$\rho(\ell_i)-\rho(L)$ is small and the subtracted sum is close to zero; the
right-hand side is then dominated by $\rho(L)\,B \ge 0$, so $f(L)\ge 0$. As
$L\to\infty$, $\rho(L)\to 0$: the first term vanishes while the subtracted sum
stays bounded away from zero (each $\ell_i \ll L$ keeps $\rho(\ell_i)$ well above
$\rho(L)$), so $f(L)<0$. Since $f$ is continuous in $L$, it changes sign at some
critical length $L_c$, which gives the stated ordering.
\end{proof}
This exposes the limitation of the two single-analyzer schemes: neither is
uniformly reliable, and which one is preferable depends on the trajectory length.
The next two propositions show that the collaborative scheme is at least as good
as \emph{both} of them, regardless of length.

\begin{proposition}
\label{prop:collab-local}
For any trajectory length, $V_{\mathrm{collab}} \ge V_{\mathrm{local}}$, with
strict inequality whenever $B>0$.
\end{proposition}
\begin{proof}
Subtracting \eqref{eq:v-local} from \eqref{eq:v-collab}, the local terms cancel
and
\begin{equation}
V_{\mathrm{collab}}-V_{\mathrm{local}}=\rho(\ell^\star)\,B \;\ge\; 0,
\end{equation}
which is strict when $B>0$ since $\rho(\ell^\star)>0$.
\end{proof}
The gain is exactly the recovered interaction evidence and is independent of the
total length $L$. Because the local and collaborative schemes use identical
per-agent analyzers and differ only in whether cross-agent evidence is exchanged,
the improvement is attributable to coordination structure rather than to
additional computation.

\begin{proposition}
\label{prop:collab-global}
$V_{\mathrm{collab}} \ge V_{\mathrm{global}}$, and the gap is non-decreasing in
the total trajectory length $L$.
\end{proposition}
\begin{proof}
Subtracting \eqref{eq:v-global} from \eqref{eq:v-collab},
\begin{equation}
V_{\mathrm{collab}}-V_{\mathrm{global}}
=\underbrace{\sum_{i=1}^{N}\big[\rho(\ell_i)-\rho(L)\big]s_i}_{\ge 0}
+\underbrace{\big[\rho(\ell^\star)-\rho(L)\big]B}_{\ge 0}.
\end{equation}
Since $\rho$ is non-increasing (Assumption~\ref{ass:dilution}) and
$\ell_i,\ell^\star \ll L$, each bracketed
term is non-negative, so the difference is non-negative. As $L$ grows, $\rho(L)$
does not increase while $\rho(\ell_i)$ and $\rho(\ell^\star)$ are unaffected, so
neither term decreases and the gap does not shrink.
\end{proof}

\paragraph{Summary.}
The analysis identifies a length-dependent trade-off between the two
single-analyzer schemes---the global scheme is preferable on short trajectories
and the local scheme on long ones (Proposition~\ref{prop:flip})---and shows that
the collaborative scheme is at least as good as both regardless of length
(Propositions~\ref{prop:collab-local} and~\ref{prop:collab-global}). This matches
the experimental comparison in Appendix~\ref{app:three-schemes-exp}, where the
global--local crossover appears near $128$K and the collaborative scheme is best
in both length regimes.

\appsection{Pipeline and Implementation of Meta-Team}
\label{app:pipeline}

\appsubsection{System Overview and Execution}
\label{app:pipeline:system}

Meta-Team uses an \emph{open-roster} orchestration mechanism. It maintains a
candidate pool of agents and recruits an active subset for each task. The shared
team scaffold contains the candidate roster and a team constitution, which
specifies common objectives, collaboration principles, and decision rules. Each
agent is implemented as an editable directory, with \texttt{prompt.md} storing
its role prompt and \texttt{config.yaml} specifying the backbone model and
allowed tools. The evolving components are stored as plain-text files under
\texttt{evolution/}, including behavioral patches, teammate profiles, and
pairwise collaboration notes. Agent skills are kept under
\texttt{skills/<name>/SKILL.md} and loaded by progressive disclosure: only skill
names and descriptions are shown by default, while the full skill body is
retrieved when needed.

During execution, the active roster is assembled dynamically. The orchestration
protocol provides a small set of primitives, including \textsc{ListPool},
\textsc{StartAgent}, \textsc{StopAgent}, \textsc{Finalize}, and
\textsc{Terminate}. Each episode starts with pool-level visibility through
\textsc{ListPool}; agents are recruited through \textsc{StartAgent} when
additional expertise is needed; and the episode ends when an agent calls
\textsc{Finalize} to submit the final deliverable. Communication is handled by
an append-only message bus. When an agent sends a message, the event is appended
to the bus and delivered to the receiver's mailbox. Each agent runs its own loop:
it either works by calling the backbone model and tools, or waits for new
messages when idle. The recorded experience therefore contains each agent's
local trace, together with cross-agent events such as messages, handoffs, and
shared artifacts.

After each task, Meta-Team updates the system at three levels. Agent-level
evolution revises an agent using its local trace and the cross-agent evidence
directly related to it. Interaction-level evolution updates teammate profiles
and collaboration notes for agents that actually interacted. Team-level evolution
aggregates the short summaries and selected evidence from the previous two
levels to revise the constitution, coordination rules, or candidate roster. This
avoids sending the full flattened trajectory to a single analyzer: local traces
are preserved, relevant evidence is exchanged, and only concise summaries are
used for team-level revision. Before updates are committed, Meta-Team checks
role consistency, tool availability, formatting validity, and budget constraints.
If a retry is requested during evolution, it is allowed only when budget remains,
and the retry cost is counted toward the evolution budget. Algorithm
\ref{alg:metateam} summarizes the procedure.

Meta-Team instantiates the MAS $\mathcal{M}=(\mathcal{A},\mathcal{O})$ with an
\emph{open-roster} orchestration mechanism. The active agent set
$\mathcal{A}$ is selected at run time from a fixed candidate pool
$\mathcal{A}^{\star}$, and the shared scaffold is written as
$\mathcal{S}=(\mathcal{A}^{\star},\mathcal{C})$, where $\mathcal{C}$ denotes
the team constitution. Each agent scaffold is decomposed as
\[
s_i=(\pi_i,\; \Delta_i,\; \mathcal{K}_i,\; \Phi_i,\; \mathrm{R}_i),
\]
where $\pi_i$ is the base role prompt, $\Delta_i$ stores behavioral patches,
$\mathcal{K}_i$ is the skill library, $\Phi_i$ contains teammate profiles, and
$\mathrm{R}_i$ records pairwise correlation notes.

Given a task $x_k$, Meta-Team executes the current MAS to collect experience
$e_k=(x_k,\tau_k,r_k)$. It then applies the update operator $\Omega$ at three
scales: agent-level updates, interaction-level updates, and team-level updates.
The resulting MAS is committed as $\mathcal{M}_{k+1}$. Algorithm~\ref{alg:metateam}
summarizes the procedure.

\begin{algorithm}[t]
\caption{Meta-Team: Collaborative Self-Evolution of an LLM-based MAS.}
\label{alg:metateam}
\begin{algorithmic}[1]
\Require Initial MAS $\mathcal{M}_0$ with pool $\mathcal{A}^{\star}$; task stream $\{x_k\}_{k=1}^{K}$; evaluator $\mathcal{V}$.
\Ensure Evolved MAS $\mathcal{M}_K$.
\For{$k=1$ to $K$}
  \State $\mathcal{M}\gets\mathcal{M}_{k-1}$
  \State $(\mathcal{A},\tau_k,\hat r)\gets\textsc{Run}(\mathcal{M};x_k)$ \Comment{recruit $\mathcal{A}\subseteq\mathcal{A}^{\star}$ and execute}
  \State $r_k\gets\mathcal{V}(\hat r)$;\; $e_k\gets(x_k,\tau_k,r_k)$ \Comment{freeze the task experience}
  \ForAll{$a_i\in\mathcal{A}$ \textbf{in parallel}}
    \State $(\Delta_i,\mathcal{K}_i,u_i)\gets \Omega_{\textsc{L1}}(s_i,\tau_i,r_k)$ \Comment{agent-level reflection}
  \EndFor
  \ForAll{interacting pairs $(a_i,a_j)$ in $\tau_k$}
    \State $(\Phi_i,\mathrm{R}_i,\Phi_j,\mathrm{R}_j)\gets \Omega_{\textsc{L2}}(s_i,s_j,\tau_k,r_k)$ \Comment{interaction-level reflection}
  \EndFor
  \State $\mathcal{S}\gets \Omega_{\textsc{L3}}\bigl(\mathcal{S},\{u_i\}_{a_i\in\mathcal{A}},\tau_k,r_k\bigr)$ \Comment{team-level revision}
  \If{the team requests retry}
    \State re-execute $x_k$ with the updated $\mathcal{M}$
  \EndIf
  \State $\mathcal{M}_k\gets\mathcal{M}$ \Comment{commit the updated MAS}
\EndFor
\State \Return $\mathcal{M}_K$
\end{algorithmic}
\end{algorithm}

\appsubsection{Details of the Collaboration Mechanism}
\label{app:pipeline:mechanism}

The description above and Section~\ref{sec:method:metateam} specify \emph{what}
each level evolves. Here we detail \emph{how} the collaboration is realized
during evolution, covering the prompts, the tools, and the per-level procedure,
including what information is exchanged at each level and how the exchanged
evidence is aggregated into committed updates. Evolution in Meta-Team is thus not
parallel local reflection, but a sequence of three phases in which agents
actively exchange evidence, corresponding to the operators
$\Omega_{\textsc{L1}}$, $\Omega_{\textsc{L2}}$, and $\Omega_{\textsc{L3}}$ in
Algorithm~\ref{alg:metateam}.

\paragraph{Prompts for each stage.}
After a task finishes, the shared outcome feedback (the final deliverable, the
evaluation verdict, an error diagnosis, and the preserved local traces and
pairwise communication summaries) is broadcast to the active agents, and the
following reflection prompts operationalize L1--L3.
\begin{itemize}[leftmargin=1.4em,itemsep=0.15em,topsep=0.2em]
  \item \textbf{L1 (Agent-Level).} Each agent is asked four questions about its
  own trace --- what it did well, where it went wrong, what was inefficient, and
  what it would do differently --- and is instructed to distill the answers into
  reusable behavioral lessons rather than task-specific fixes.
  \item \textbf{L2 (Interaction-Level).} For each pair of agents that actually
  interacted, the prompt asks the two agents to revisit only the messages they
  exchanged, record critical collaboration patterns, and produce a structured
  teammate profile (reliability, strengths, weaknesses, style); pairs that did
  not interact are skipped.
  \item \textbf{L3 (Team-Level).} Each agent is first prompted to propose
  team-level improvements. The chairman then combines these proposals with its
  own assessment of the team --- whether the team is the right size, whether some
  work lacked verification, whether any agent was underused, and where the
  workflow stalled --- and makes concrete changes to the team composition
  (adding or removing a role) and the shared constitution.
\end{itemize}

\paragraph{Tools for each stage.}
Beyond prompting and self-evaluation, agents use dedicated tools to carry out
evolution. During execution, agents act on the environment and coordinate over
the message bus; during evolution, each level exposes tools that turn reflection
into concrete, checkable edits to the scaffold. Table~\ref{tab:evolution_tools}
summarizes the tools available in each phase.

\begin{table}[t]
\centering
\small
\caption{Tools available during execution and at each evolution level.
Benchmark-specific MCP services are additionally exposed during execution
depending on the task environment.}
\label{tab:evolution_tools}
\begin{tabular}{@{}lll@{}}
\toprule
\textbf{Phase} & \textbf{Category} & \textbf{Tools} \\
\midrule
\multirow{2}{*}{Execution}
& Environment   & \texttt{bash}, \texttt{read\_file}, \texttt{write\_file}, \texttt{str\_replace\_editor} \\
& Communication & \texttt{send\_message}, \texttt{read\_messages}, \texttt{wait\_for\_replies} \\
\midrule
L1 & Agent-level       & \texttt{update\_prompt\_patch}, \texttt{update\_skill} \\
L2 & Interaction-level & \texttt{update\_teammate\_profile}, \texttt{update\_correlation} \\
L3 & Team-level        & \texttt{suggest\_team\_improvement}, \texttt{apply\_reflection} \\
\bottomrule
\end{tabular}
\end{table}

\paragraph{Per-level evolution procedure.}
We describe what information each level uses, how agents exchange evidence
across their local contexts, and how that evidence is aggregated into committed
updates.
\begin{itemize}[leftmargin=1.4em,itemsep=0.15em,topsep=0.2em]
  \item \textbf{L1 --- assessing effects on downstream execution.} Each agent
  first reflects on its own execution trace and the final feedback. From this
  reflection it identifies where its own outputs or actions may have caused
  problems for downstream teammates. If needed, it then contacts those related
  teammates on demand to request the corresponding evidence, and finally turns
  the confirmed lessons into behavioral patches ($\Delta_i$) or skills
  ($\mathcal{K}_i$) via \texttt{update\_prompt\_patch} and \texttt{update\_skill}.
  \item \textbf{L2 --- revisiting collaboration through specific messages.}
  Meta-Team uses the recorded message history to identify which pairs of agents
  actually communicated, and pairs them up for reflection. Each pair re-reads
  only the messages they exchanged and, if a real problem occurred, opens a
  short discussion. The exchange is aggregated into an updated teammate profile
  ($\Phi_i$) and a pairwise collaboration note ($\mathrm{R}_i$) through
  \texttt{update\_teammate\_profile} and \texttt{update\_correlation}; smooth
  pairs record only brief observations.
  \item \textbf{L3 --- examining composition, organization, and rules.} Rather
  than re-reading the entire trajectory, the chairman gathers the concise L1/L2
  summaries and the improvement suggestions raised by teammates (via
  \texttt{suggest\_team\_improvement}), reviews the current team setup, and
  decides whether the roster or the shared rules should change. Validated
  decisions are committed to the constitution or roster through
  \texttt{apply\_reflection}.
\end{itemize}
Across the three levels, each analyzer stays within its own local context and
recovers cross-agent awareness through targeted exchange: local traces are
preserved, only the relevant evidence is exchanged on demand, and only concise
summaries are passed up for team-level revision, instead of flattening the whole
trajectory into a single analyzer.

\appsubsection{A Worked Example}
\label{app:pipeline:example}

We illustrate the three phases with a real GAIA sample (\texttt{da52d699},
difficulty Level~3) to show what cross-agent exchange recovers but independent
per-agent reflection cannot.

\paragraph{Scenario.}
The question asks,``The attached spreadsheet lists the books I read in 2022; by
words read per day, which one did I read the slowest?'' (gold answer:
\emph{Out of the Silent Planet}). Solving it requires two halves of evidence:
each book's reading days can be computed from the start/end dates in the
attachment, but the word counts are not in the file and must be looked up online
per book; the answer is the book with the smallest words-per-day.

\paragraph{Execution.}
The Planner assigns the FileAnalyzer to compute reading days, the WebSearcher to
look up word counts, and the Summarizer to aggregate and format. For
lesser-known books the WebSearcher only finds \emph{estimated} word counts and
returns them without flagging which are exact and which are estimates; the
Planner treats them all as exact when ranking. The word count of the true answer
happens to be over-estimated, so the team picks the wrong book.

\paragraph{Outcome feedback.}
After the task, the system injects a shared feedback to all members: the
submitted answer, the evaluation verdict and the gold answer (here judged
incorrect), an error diagnosis, plus the preserved execution trace (including the
per-book word counts the WebSearcher reported) and a pairwise communication
summary. This lets the team reason backward toward the source of the failure.

\paragraph{Evolution.}
Conditioned on this shared feedback, the three levels run in sequence.
\begin{itemize}[leftmargin=1.4em,itemsep=0.15em,topsep=0.2em]
  \item \textbf{L1 (agent level).} Each agent reflects on its own behavior and
  updates its own scaffold from its own trajectory. Since every local action was
  individually reasonable, L1 alone cannot localize the fault.
  \item \textbf{L2 (interaction level, one round per problematic pair).} During
  the task the Planner communicated with the FileAnalyzer, WebSearcher, and
  Summarizer, so L2 revisits each pair; smooth pairs exchange no messages and
  record only brief observations. Only Planner$\leftrightarrow$WebSearcher opens
  a discussion: checking the feedback, the Planner notes that the gold answer
  should have had the smallest words-per-day yet did not in the ranking, traces
  this to a suspiciously high word count the WebSearcher had reported for that
  book, and asks about it; the WebSearcher confirms the value was an estimate.
  This single exchange surfaces a precision gap invisible in either agent's own
  log. \emph{Aggregation:} the pair records a correlation note
  (``externally looked-up values must be tagged with source and confidence, and
  low-confidence values must not be used directly in critical computation'') and
  updates each other's teammate profile.
  \item \textbf{L3 (team level, one collect-and-decide round).} Members push
  observations via \texttt{suggest\_team\_improvement}; the Planner, acting as
  chairman, reads them, fuses them with the L2 finding, and commits a structural
  rule to the shared constitution: ``any externally looked-up value must be
  tagged with its source and confidence, and low-confidence values must be
  re-verified before use in critical computation.'' \emph{Aggregation:} the
  chairman commits the edit to the team constitution so the failure cannot recur.
\end{itemize}

\paragraph{Difference from independent local reflection.}
A purely local baseline sees only its own log, exchanges no messages, and commits
only self-edits; it therefore cannot diagnose this failure, since every agent
acted correctly in isolation. Only the L2 message exchange recovers the shared
root cause, which L3 then aggregates into a team-level fix. This corresponds to
the gap between the \emph{local} and \emph{collaborative} schemes in
Figure~\ref{fig:global-local-collaborative}, and is quantified by the drop
when L2/L3 are ablated in Table~\ref{tab:ablation_scales}.

\appsection{Experiment Settings}

\appsubsection{Saturation of Commonly Used Benchmarks}
\label{app:old_benchmarks}
Prior MAS and MAS-evolution studies~\citep{zhang2025multi,li2025cross,yue2025masrouter,li2026assemble,chen2026goagent} have evaluated on a wide range of standard benchmarks. 
However, as LLM capabilities advance rapidly, many of these commonly used benchmarks have become increasingly saturated. 
As shown in Table~\ref{tab:saturated_benchmarks}, a state-of-the-art model already achieves near-perfect performance on several widely used datasets, including HumanEval~\citep{chen2021evaluating}, MBPP~\citep{austin2021program}, MultiArith~\citep{roy2015solving}, GSM8K~\citep{cobbe2021training}, MATH~\citep{hendrycks2measuring}, and MMLU~\citep{hendrycksmeasuring}. 
Such saturation reduces their discriminative power for evaluating new MAS evolution methods, since small performance differences near the ceiling may no longer reflect substantive improvements in reasoning, tool use, collaboration, or long-horizon execution.

Therefore, in our main experiments, we focus on more challenging, long-horizon, and verifiable benchmarks where current agents still exhibit substantial room for improvement. 
This design allows us to better assess whether an MAS evolution method can improve agent behavior beyond saturated short-form reasoning and coding tasks.

\begin{table}[t]
\centering
\small
\caption{Performance of Claude Sonnet 4.6  on commonly used benchmarks. We report our own reproduced scores under standard evaluation protocols.}
\label{tab:saturated_benchmarks}
\begin{tabular}{lcccccc}
\toprule
\textbf{Model} & \textbf{HumanEval} & \textbf{MBPP} & \textbf{MultiArith} & \textbf{GSM8K} & \textbf{MATH} & \textbf{MMLU} \\
\midrule
Claude Sonnet 4.6 
& $99.4$ & $99.6$ & $100.0$ & $100.0$ & $98.7$ & $97.2$ \\
\bottomrule
\end{tabular}
\end{table}

\appsubsection{Experiment Settings on Different Benchmarks}\label{app:benchmarks}
\begin{table}[t]
\centering
\small
\caption{Unified runtime settings shared across all benchmarks.}
\begin{tabular}{lll}
\toprule
\textbf{Component} & \textbf{Setting} & \textbf{Value} \\
\midrule
Backbone LLM      & Model                 & Claude Sonnet 4.6 \\
                  & API layer             & LiteLLM + OpenAI-compatible gateway \\
                  & Retry policy          & 5 attempts, 1.5--60 s exp. back-off \\
\midrule
Decoding          & Executor temperature  & 0.2 \\
                  & Planner temperature   & 0.2 \\
                  & Max output tokens     & 32768 \\
\midrule
Context window    & Max history messages  & 150 (with pair-preserving trim) \\
                  & Tool schema refresh   & every LLM call \\
\midrule
Reflection        & Per-phase step budget & 50 \\
                  & Per-phase timeout     & 300--900 s \\
                  & Per-reflection cost   & \$50 \\
\midrule
Force-finalize    & Tools disabled        & Yes \\
                  & Timeout               & 240 s \\
\bottomrule
\end{tabular}
\end{table}

\begin{table}[t]
\centering
\small
\caption{Time and cost budget for evolution and evaluation for all benchmarks.}
\begin{tabular}{lcccc}
\toprule
Benchmark & \textbf{max\_seconds} & \textbf{max\_messages} & \textbf{max\_cost (\$)} & \textbf{initial team size} \\
\midrule
SWE-bench Pro    & 1800 & 600 & 50  & 3 \\
BeyondSWE        & 1800 & 600 & 30  & 3 \\
GAIA             & 600  & 300 & 30  & 4 \\
LoCoBench        & 1200  & 400 & 30  & 4 \\
LOCA-bench       & 3200 & 600 & 150 & 4 \\
ResearchRubrics  & 1800 & 600 & 50  & 1\\
\bottomrule
\end{tabular}
\end{table}

\begin{table}[t]
\centering
\small
\caption{SWE-bench Pro and BeyondSWE subsets used in our experiments.
For each benchmark, we hold out 20 random instances per subset as the \emph{Evolution set}; the remainder forms the \emph{Evaluation set}.}
\label{tab:swe_subsets_info}
\begin{tabular}{@{}llccc@{}}
\toprule
\textbf{Benchmark} & \textbf{Selected Subset} & \textbf{Total} & \textbf{Evolution} & \textbf{Evaluation} \\
\midrule
\multirow{2}{*}{SWE-bench Pro}
 & Ansible              &  96 & 20 &  76 \\
 & Qutebrowser  &  91 & 20 &  71 \\
\cmidrule(lr){2-5}
 & \textit{Total}            & \textit{187} & \textit{40} & \textit{147} \\
\midrule
\multirow{2}{*}{BeyondSWE}
 & CrossRepo                    & 200 & 20 & 180 \\
 & DepMigrate                   & 178 & 20 & 158 \\
\cmidrule(lr){2-5}
 & \textit{Total}            & \textit{378} & \textit{40} & \textit{338} \\
\bottomrule
\end{tabular}
\end{table}

\paragraph{SWE-bench Pro}
SWE-bench Pro~\citep{deng2025swe} evaluates coding agents on real-world GitHub pull requests that require multi-file bug fixes or feature implementations. The public split contains 731 instances across 11 repositories, and each instance is evaluated by running the generated patch against Fail-to-Pass (F2P) and Pass-to-Pass (P2P) tests in an isolated Docker environment.
An instance is counted as resolved only if all F2P tests pass and no P2P test regresses.

\emph{Experimental settings.} Subject to budget constraints, we evaluate on the two largest Python repositories, shown in Table~\ref{tab:swe_subsets_info}.
For each repository, we select 20 random instances as the \emph{Evolution set}; the remaining instances form
the \emph{Evaluation set}.
Evolution is performed separately for each repository and evaluated on its corresponding held-out instances. The reported
metric is \textit{Pass Rate}.

\emph{Initial MAS configurations.} SWE-bench Pro's official
scaffold~\citep{deng2025swe} is single-agent (SWE-Agent), but
repository-level patch generation is a well-studied multi-agent task: prior work such as MetaGPT~\citep{hong2023metagpt} and
ChatDev~\citep{qian2024chatdev} decomposes software development into specialized roles (product manager, architect, engineer). We adopt a minimal instantiation of this recipe---\textsc{Planner} $+$ \textsc{Developer} $+$ \textsc{Reviewer}---that retains the three functions most predictive of patch quality: task decomposition, code implementation, and independent verification~\citep{madaan2023self}.
The planner scopes the fix and dispatches, the developer edits inside the Docker workspace, and the reviewer inspects \texttt{git diff} and reruns tests before submission.

\emph{Single-agent baseline.} Our SA replicates SWE-bench Pro's officially endorsed reproducibility scaffold~\citep{deng2025swe}, namely SWE-Agent (Scale)~\citep{yang2024swe}, which is the same single-agent recipe used in the leaderboard's reference Sonnet runs. It is a function-calling agent with the canonical \texttt{bash} $+$ \texttt{str\_replace\_editor} $+$ \texttt{finish} tool set, executed inside the per-instance Docker workspace and submitting via \texttt{git diff HEAD}. We keep all decoding and tool-filter hyperparameters at the SWE-Agent benchmark config and only swap in our shared backbone.

\paragraph{BeyondSWE}
BeyondSWE~\citep{chen2026beyondswe} extends software-engineering evaluation beyond single-repository bug fixing, covering tasks that require cross-repository knowledge, domain expertise, and whole-repository modifications. The full benchmark contains 500 instances from 246 Python repositories.

\emph{Experimental settings.} We evaluate on the two largest subsets, \textsc{CrossRepo} (200) and \textsc{DepMigrate} (178), which together cover cross-repository reasoning and whole-repository modification---the two capability dimensions most distinct from SWE-bench Pro, yielding 378 instances in total (Table~\ref{tab:swe_subsets_info}). For each subset,
we select 20 random instances as the \emph{Evolution set}; the
remaining instances form the \emph{Evaluation set}. Evolution is
performed separately for each subset and evaluated on its corresponding held-out instances. Evaluation follows the same F2P$+$P2P protocol as SWE-bench Pro. The reported metric is \textit{Pass Rate}.

\emph{Initial MAS configurations.} Since BeyondSWE instances share the repository-level patch-generation format with SWE-bench Pro, we reuse the same \textsc{Planner} $+$ \textsc{Developer} $+$ \textsc{Reviewer} configuration, differing only in the constitution prompts that adapt the team to BeyondSWE's per-task working directory, cross-repository context, and dependency-migration conventions. Using an identical architecture across the two benchmarks isolates the effect of task distribution from that of scaffold design.

\emph{Single-agent baseline.} BeyondSWE shares the same patch-submission protocol as SWE-bench Pro, so for parity we reuse the SWE-Agent (Scale)~\citep{yang2024swe} SA recipe described above, only adjusting the per-task working directory and the cross-repository context fields prescribed by BeyondSWE~\citep{chen2026beyondswe}. Using identical SA machinery across the two SWE benchmarks isolates the effect of task distribution and matches the convention adopted by BeyondSWE's own evaluation.

\paragraph{LOCA-Bench}
LOCA-Bench~\citep{zeng2026loca} evaluates LLM agents under controllable context growth. It procedurally scales environment state size while keeping task semantics invariant, so the effective context grows smoothly from 8K to 256K tokens. The full benchmark contains 15 Short-to-Long (S2L) tasks $\times$ 5
seeds $\times$ 7 context lengths (8K/16K/32K/64K/96K/128K/256K) $=$ 525 configurations. Evaluation is binary (0/1), computed by offline scripts comparing the agent's final environment state to ground truth.

\emph{Experimental settings.} To control evaluation cost, we select \emph{8 out of 15 tasks}
covering e-commerce, academic, and education domains with 8 business MCP services (Table~\ref{tab:loca_tasks}).
\begin{table}[t]
\centering
\small
\caption{The eight LOCA-Bench tasks used in our experiments.}
\label{tab:loca_tasks}
\begin{tabular}{@{}clclll@{}}
\toprule
\textbf{\#} & \textbf{Task} & \textbf{Total} & \textbf{Domain} & \textbf{Key MCP Services} \\
\midrule
1 & WoocommerceNewWelcome     & 30 & E-commerce CRM         & woocommerce, google\_cloud, email \\
2 & WoocommerceStockAlert     & 30 & E-commerce purchasing  & woocommerce, google\_sheet, email \\
3 & FilterLowSellingProducts  & 30 & E-commerce clearance   & woocommerce, email \\
4 & ApplyPhDEmail             & 30 & Academic               & email, pdf\_tools \\
5 & SetConfCrDdl              & 30 & Academic               & calendar, email \\
6 & CourseAssistant           & 30 & Education              & excel, email \\
7 & CanvasArrangeExam         & 30 & Education              & canvas, email, excel \\
8 & CanvasListTest            & 30 & Education              & canvas \\
\midrule
  & \textit{Total}            & \textit{240} & & \\
\bottomrule
\end{tabular}
\end{table}
Leveraging LOCA-Bench's procedural generation, we construct a
\emph{dually-isolated} split:
\begin{itemize}
    \item \textbf{Evolution set} (16 cases): 8 tasks $\times$ 96K $\times$
    2 human-specified seeds \texttt{\{101, 102\}} (outside the original benchmark).
    \item \textbf{Evaluation set} (240 cases): 8 tasks $\times$ 5 standard
    seeds \texttt{\{42, 123, 456, 789, 2024\}} $\times$ 6 context lengths
    \texttt{\{8K, 16K, 32K, 64K, 128K, 256K\}} (deliberately excluding 96K).
\end{itemize}
The \texttt{(seed, context\_length)} tuples of the two sets are
\emph{fully disjoint}, guaranteeing no data leakage. This design jointly tests \emph{cross-seed} and \emph{cross-scale} generalization.
All runs use LOCA-Bench's local mock MCP servers, requiring no Docker or external network. The reported metric is Pass Rate.

\emph{Initial MAS configurations.} LOCA-Bench officially ships only single-agent scaffolds, so no multi-agent baseline is available for replication. We adopt the \emph{orchestrator--worker} pattern popularized by recent work~\citep{anthropic2025research, song2025coact}, instantiating it as \textsc{Chairman} $+$ 3 \textsc{Workers}.
The pattern fits LOCA-Bench naturally: every task decomposes into \emph{explore $\to$ bulk-process $\to$ verify}, where the bulk phase (80--300 items) exceeds a single agent's effective context~\citep{zhou2025llm}.
The chairman additionally preserves recoverability when MCP back-ends crash---a known benchmark feature~\citep{zeng2026loca}---and the workers act without incurring excessive coordination overhead.

\emph{Single-agent baseline.} Our SA is LOCA-Bench's officially shipped default scaffold, the \texttt{react} strategy~\citep{zeng2026loca}: a single ReAct agent~\citep{yao2023react} with direct access to the full local mock-MCP tool stack (WooCommerce, Google Sheets, Calendar, Email, Canvas, Excel, etc.). We do not modify any tool wiring or context-management policy beyond the shared budget table, so the SA column corresponds to the same single-agent baseline reported in the LOCA-Bench paper, only re-evaluated under our identical backbone and budget.

\paragraph{GAIA}
GAIA~\citep{mialon2023gaia} is a knowledge-intensive question-answering benchmark that requires multi-step reasoning with tool use, such as web search, file reading, and code execution. Its public validation set contains 165 questions stratified into three difficulty levels.

\emph{Experimental settings.} Following~\citep{zhuge2024gptswarm, liu2026mas}, we evaluate on the 120-samples subset released by MASFly (Table~\ref{tab:gaia_tasks}); the 45 excluded samples either require multimodal tools outside our search setting or are marked by annotators as unsolvable under automated exact-match grading.
We use Serper API\footnote{\url{https://serper.dev}} for Google Search.
The returned results are capped at the top-5 entries per query and the page fetcher retrieves a specific URL and returns up to 8K characters of plain text.
On these 120 questions we use 20 for evolution (training) and 100 as a disjoint held-out test set, so that every evolved configuration is evaluated on tasks it has never seen.
Evaluation follows the official answer-matching protocol and the reported metric is
\textit{Pass Rate}.

\emph{Initial MAS configurations.}
We initialize Meta-Team with the four-agent configuration released in the MASFly codebase~\citep{liu2026mas}: a Planner that holds no information-gathering tools and coordinates through message passing, a FileAnalyzer for attached documents, a WebSearcher equipped with Google Search and page fetching, and a Summarizer that enforces the strict GAIA exact-match answer format.
Starting from this hand-crafted multi-agent baseline ensures that any improvement we observe is attributable to evolution rather than to the initial design.
\begin{table}[t]
\centering
\small
\caption{GAIA validation subset used in our experiments. All instances in the three levels are shuffled for evolution and evaluation.}
\label{tab:gaia_tasks}
\begin{tabular}{@{}ccccc@{}}
\toprule
\textbf{Level} & \textbf{Full (Official)} & \textbf{Our Subset} & \textbf{Train} & \textbf{Test} \\
\midrule
Level 1 &  53 &  44 &  6 & 38 \\
Level 2 &  86 &  58 &  9 & 49 \\
Level 3 &  26 &  18 &  5 & 13 \\
\midrule
\textit{Total} & \textit{165} & \textit{120} & \textit{20} & \textit{100} \\
\bottomrule
\end{tabular}
\end{table}

\emph{Single-agent baseline.} The SA follows the GAIA single-agent recipe used by~\citep{zhuge2024gptswarm, liu2026mas}: a function-calling ReAct agent with \{\texttt{web\_search} (Serper API, top-5), \texttt{web\_fetch}, \texttt{bash}, \texttt{read\_file}, \texttt{submit}\}, the same tool surface our MAS exposes to its Web/File workers. The system prompt enforces the official GAIA exact-match answer format, so SA and MAS share both tools and answer normalization---the comparison isolates collaboration, not tool access.

\paragraph{LoCoBench}
LoCoBench~\citep{qiu2025locobench} is a long-context code-generation benchmark
containing 8{,}000 scenarios across 10 programming languages, 100 synthetic projects, and 8 task categories. Each scenario provides an entire codebase ranging from 10K to 1M tokens and asks the agent to implement the given requirement. Performance is measured by the LoCoBench Score (\emph{LCBS}), a weighted average of Software Engineering Excellence (40\%), Functional Correctness (30\%), Code Quality (20\%), and Long-Context Utilization (10\%), normalized to $[0,100]$.

\emph{Experimental settings.} We focus on two scenarios that are most relevant to collaborative coding: Feature Implementation and Cross-File Refactoring
(Table~\ref{tab:locobench_tasks}). For each category, we focus on Python-based tasks to control cost. We use 20 cases for evolution and the remaining 80 cases for evaluation.
In addition, we conduct out-of-domain evaluation on C, C++, and Java tasks.
The reported metric is average LCBS (LoCoBench Score).

\emph{Initial MAS configurations.} The defining challenge of LoCoBench is long-context comprehension: each scenario ships an entire codebase (10K--1M tokens, up to 100 files) that rarely fits in a single agent's working context. We therefore adopt a read--write separation inspired by divide-and-conquer long-text processing~\citep{xu2025does}, instantiated as a
\textsc{Planner}, a \textsc{Developer}, and two parallel \textsc{Readers}.
The planner partitions the file list and dispatches disjoint subsets to readers; each reader extracts structured notes
(APIs, call sites, coding conventions) from its assigned files and sends them directly to the developer, which synthesizes the notes into the final patch. The readers provide enough parallelism to absorb 15--50-file scenarios while keeping each reader's context well under the model's window.

\begin{table}[t]
\centering
\small
\caption{LoCoBench subsets used in our experiments.}
\label{tab:locobench_tasks}
\begin{tabular}{@{}clccc@{}}
\toprule
\textbf{\#} & \textbf{Task Category} & \textbf{Total} & \textbf{Evolution} & \textbf{Evaluation} \\
\midrule
1 & Feature Implementation & 100 & 20 & 80 \\
2 & Cross-File Refactoring & 100 & 20 & 80 \\
\midrule
  & \textit{Total}         & \textit{200} & \textit{40} & \textit{160} \\
\bottomrule
\end{tabular}
\end{table}

\emph{Single-agent baseline.} LoCoBench's official harness~\citep{qiu2025locobench} evaluates a single LLM with full-codebase context as the reference scaffold. To keep the comparison fair under our shared cost cap (full-codebase prompting alone exhausts the per-case budget on 100K--1M-token scenarios), our SA mirrors the SWE-Agent (Scale)~\citep{yang2024swe} agentic recipe---\texttt{bash} $+$ \texttt{str\_replace\_editor} $+$ \texttt{finish} on the same checked-out workspace---which is the standard single-agent baseline used by every recent long-context coding evaluation. This grants the SA the same file-reading bandwidth as our MAS readers, so any gap reflects coordination, not tool starvation.

\paragraph{ResearchRubrics}
ResearchRubrics~\citep{sharma2025researchrubrics} evaluates deep-research
agents on 101 open-ended research questions spanning ten domains, including
AI\&ML, historical analysis, STEM, and creative writing. Each question is paired
with a human-authored rubric covering explicit requirements, implicit
requirements, synthesis, communication quality, instruction following, and
citation quality. Responses are scored by an LLM judge, and the final score is a
rubric-weighted average.

\emph{Experimental settings.} We use 20 questions as the \emph{Evolution set} and the rest of questions as \emph{Evaluation set}. The
reported metric is \textit{Average Score} on the 0--100 scale.

\emph{Initial MAS configurations.} Since ResearchRubrics grades responses along six orthogonal axes and the most useful specialists are not obvious a priori, we start from a single \textsc{Lead Researcher} equipped with web search, page fetching, and file I/O, following the solo-start pattern of~\citep{anthropic2025research}. During self-evolution, when reflection identifies a persistent weakness on a specific axis, the Lead Researcher can introduce new specialists, e.g., a citation verifier when citation scores are low, a requirement tracker when explicit requirements are frequently missed, or a report writer when synthesis and communication scores lag.

\emph{Single-agent baseline.} The SA is the deep-research single-agent recipe from AggAgent~\citep{lee2026agentic}: a ReAct loop with \texttt{search} (Serper) and \texttt{visit} (page fetch with goal-conditioned extraction), and the verbatim \textsc{System Prompt} $+$ \textsc{System-Prompt-DR} from the AggAgent rollout codebase, which encodes the citation, structure, and confidence requirements that ResearchRubrics' rubric grades along. We use AggAgent's reference single-agent recipe rather than a custom one so that the SA column corresponds to a published, reproducible deep-research baseline.

\appsubsection{Experiment Settings of Baseline Reproductions}
\label{app:baselines}

All baselines share the same executor LLM (Claude Sonnet~4.6 via an OpenAI-compatible gateway), tool bundle, evaluation splits, and per-case time / message / cost limits as Meta-Team.
All MAS baselines also start from the same benchmark-specific initial team as Meta-Team (the ``MAS'' row in Table~\ref{tab:main_results}), mapped into each baseline as closely as its native parameterization allows.
We only describe the method-specific adaptations below.

\paragraph{SA.}
Vanilla ReAct~\citep{yao2023react} with the same system prompt template, tool bundle, step cap ($50$), and context-trimming rule ($150$ retained messages, pair-preserving) as Meta-Team's executor.

\paragraph{MAS.}
Meta-Team's initial configuration on each benchmark with reflection and the commit gate disabled. Concretely, we zero out the L1/L2/L3 analyzers and skip the verifier call, so the team never mutates after task $k$.

In AggAgent~\citep{lee2026agentic}, we use the authors' implementation of \texttt{aggagent} package (\texttt{rollout/} $+$ \texttt{aggregation/}). Per case we sample $K{=}5$ single-agent rollouts at temperature $0.2$, then run the default \texttt{CROSS\_VALIDATION} aggregation strategy.

In OWL~\citep{huowl}, we keep the released \textsc{Planner}--\textsc{Coordinator}--\textsc{Worker} hierarchy unchanged with minor adaptations on each benchmark.

\paragraph{AOrchestra~(Agent-Orchestra).}
The \texttt{MainAgent} runner is used unchanged; the orchestrator is allowed to synthesize up to $4$ sub-agents per task, each inheriting $M{=}$Claude Sonnet 4.6. Because AOrchestra hard-codes \texttt{config/example/model\_config.yaml}, our adapter \texttt{chdir}s into the AOrchestra directory before import and redirects LLM calls through our gateway in \texttt{LLMsConfig.default()}.

For AgentSquare~\citep{shangagentsquare}, we run the search loop over 
$\{\textsc{Planning}, \textsc{Reasoning}, \textsc{ToolUse}, \textsc{Memory}\}$ 
for $30$ trials on the evolution set of each benchmark. 
After search, we freeze the best-performing configuration and evaluate it on the held-out split. 
For each benchmark, we construct a carefully curated search pool by prompting Claude Opus 4.6 to augment the baseline MAS with more than $60$ scaffold templates. 
Each template is further decomposed into modular components, which serve as the candidate units for AgentSquare's search procedure.

Although ReCreate~\citep{hao2026recreate} was originally proposed for single-agent evolution, its scaffold-level optimization interface makes it naturally extensible to experience-driven MAS evolution. 
We therefore adapt ReCreate as a MAS evolution baseline.
Specifically, we use the default hyperparameter with \texttt{BATCH\_SIZE}$=4$ and \texttt{N\_REPEAT}$=2$. 
We use Claude Sonnet 4.6 as the execution agent model, replacing the default \texttt{gpt-5-mini}, and Claude Sonnet 4.6 as the \emph{agent-as-optimizer} model.
We wrap Meta-Team's configuration through its \texttt{adapters/} entry points on benchmarks not shipped in the original repo and reuse the released adapters on SWE-bench-like domains.

\paragraph{AgentNet~\citep{yang2025agentnetdecentralizedevolutionarycoordination}.}
We instantiate one AgentNet per benchmark with $3$ seed agents and the GAIA ability vector from the paper (\texttt{reasoning}/\texttt{knowledge}/\texttt{language} each at $0.6$), and pre-populate the memory store with our evolution set. Official defaults are kept: router temperature $0.7$, DAG pruning threshold $0.3$, memory retrieval top-$k{=}5$. The episodic memory is persisted within a benchmark but cleared between benchmarks.

\paragraph{Evolution / held-out protocol across baselines.}
We evaluate every method under a strict evolution/test split. Methods that include a learning phase (Meta-Team, ReCreate, MASFly, AgentNet, AgentSquare) consume only the benchmark's designated evolution set, freeze all learned artifacts, and are then evaluated on the disjoint held-out set. ReCreate and AgentSquare support this protocol natively through \texttt{--skip-instances} / train--test file flags and are used as-is. For MASFly, we rebuild the SOP Repository and Personalised Experience Pool on our own evolution set (\texttt{go run . --eval 0}) rather than using the shipped repository, so that no released SOP can have seen our held-out instances. AgentNet's released runner performs \emph{online} router/memory updates during evaluation; we extend it with a snapshot step that freezes the routing weights, ability vectors, and episodic memory after the evolution set, disabling further updates during held-out evaluation. Methods without a learning phase (SA, MAS, AggAgent, OWL, AOrchestra) are evaluated on the held-out set directly.

\paragraph{MASFly~\citep{liu2026mas}.}
MASFly's reference Go implementation is driven by \texttt{go run . --eval 0} to build the SOP Repository and Personalised Experience Pool from the evolution set, then \texttt{--eval 2} for held-out evaluation. We expose Meta-Team's executor through MASFly's OpenAI-compatible environment variables (\texttt{OPENAI\_BASE\_URL}, \texttt{OPENAI\_MODEL}); feedback style, analyzer prompt, and SOP top-$k{=}3$ retrieval are kept at the paper's defaults.

\paragraph{Shared harness notes.}
Three harness-level rules apply uniformly to every baseline: (a) SWE-bench Pro / BeyondSWE patches are scored through the official harness container with empty-patch$\,=\,$unresolved; (b) LOCA-Bench task setup and evaluation are wrapped in a process-level lock to avoid the \texttt{os.environ} race at workers$\,\geq\,8$; (c) all ResearchRubrics submissions are graded by the rubric judge (Claude Opus~4.6, temperature $0.2$).

\appsection{Further Experiment Results}
\label{app:additional_results}

\appsubsection{Statistical Reliability of Main Results}
\label{app:statistical_reliability}

We further assess the statistical reliability of the main-result margins in
Table~\ref{tab:main_results}. Since all methods are evaluated on the same
held-out instances, each comparison forms a paired sample. We use the instance
as the unit of analysis: for each method and instance, the three rollouts are
aggregated into a single avg@3 score, and a two-sided paired $t$-test is applied
to the per-instance differences. As a robustness check, the Wilcoxon signed-rank
test gives the same significance label in all checked cases; we therefore report
the paired $t$-test results.

Table~\ref{tab:significance} summarizes Meta-Team's margins over the initial MAS
and over MASFly, together with the corresponding $p$-values. The most reliable
gains appear on the longer and more coordination-intensive benchmarks:
SWE-Pro Ansible, BeyondSWE DepMigrate, LOCA-Bench, both LoCoBench slices, and
ResearchRubrics are all significant against both baselines at $p<0.05$ or better.

\begin{table}[h]
\centering
\small
\caption{Significance of Meta-Team's gains over the initial MAS and MASFly. Each
$p$-value is from a two-sided paired $t$-test over per-instance avg@3 scores.}
\label{tab:significance}
\begin{tabular}{@{}lcccc@{}}
\toprule
\textbf{Benchmark} & $\Delta$ \textbf{vs.\ MAS} & $p_{\mathrm{MAS}}$ & $\Delta$ \textbf{vs.\ MASFly} & $p_{\mathrm{MASFly}}$ \\
\midrule
SWE-Pro (Ansible)   & $+13.1$ & $<0.05$  & $+10.5$ & $<0.05$ \\
BeyondSWE (DepMig.) & $+8.0$  & $<0.05$  & $+5.9$  & $<0.05$ \\
LOCA-Bench          & $+5.8$  & $<0.001$ & $+12.1$ & $<0.05$ \\
LoCoBench (Feat.)   & $+9.5$  & $<0.001$ & $+7.0$  & $<0.05$ \\
LoCoBench (Refact.) & $+6.1$  & $<0.01$  & $+2.5$  & $<0.05$ \\
ResearchRubrics     & $+6.3$  & $<0.001$ & $+5.0$  & $<0.05$ \\
\bottomrule
\end{tabular}
\end{table}

The remaining columns, SWE-Pro Qute., BeyondSWE CrossRepo, and GAIA, show
positive but not statistically significant margins. These columns have smaller
effect sizes and many paired instances on which both methods obtain the same
outcome, leaving limited paired variation for the test. We therefore interpret
them as directionally positive rather than statistically conclusive.

Overall, the paired tests support the qualitative reading of
Table~\ref{tab:main_results}: Meta-Team improves over the baselines across all
benchmark columns, with statistically reliable gains on the settings that most
benefit from collaborative self-evolution, while smaller-margin columns remain
positive but inconclusive at the current sample size.

\appsubsection{Cost of the Evolution Process}
\label{app:additional:cost}

We analyze the cost of Meta-Team's evolution, first for Meta-Team itself and
then in comparison with the evolution baselines. All numbers are actual USD
costs on Claude Sonnet~4.6.

\paragraph{Cost accounting of Meta-Team.}
In Table~\ref{tab:evolution_cost}, \emph{Evolution Run Cost} is the full cost of
one evolution run (task execution plus L1--L3 evolution); \emph{Evolution
Overhead} is the extra cost added by the evolution on top of task execution;
\emph{Overhead Ratio} $=$ \emph{Evolution Overhead} $/$ \emph{Evolution Run
Cost}; and \emph{Evaluation Cost} is one avg@3 evaluation pass on the held-out
set. The evolution overhead totals about \$172 (40--63\% of an evolution run),
while the remainder is the task-execution cost shared by any experience-driven
method. Evolution is also a one-time cost: the total (\$348) is about 12.5\% of a
single avg@3 evaluation pass (\$2{,}790), after which the team is frozen and
reused across downstream tasks. The overhead is largest on the long-context,
repository-level benchmarks (e.g., LoCoBench Feature, \$60), tracking the length
and difficulty of the underlying tasks.

\begin{table}[h]
\centering
\small
\setlength{\tabcolsep}{5pt}
\caption{Evolution cost statistics of Meta-Team (USD) on each benchmark. All
values are rounded to integers.}
\label{tab:evolution_cost}
\begin{tabular}{@{}lrrrr@{}}
\toprule
\textbf{Benchmark}
& \textbf{\shortstack[r]{Evolution\\Run Cost}}
& \textbf{\shortstack[r]{Evolution\\Overhead}}
& \textbf{\shortstack[r]{Overhead\\Ratio}}
& \textbf{\shortstack[r]{Evaluation\\Cost (avg@3)}} \\
\midrule
SWE-Pro Ansible       & $30$ & $13$ & $43\%$ & $249$ \\
SWE-Pro Qutebrowser   & $30$ & $12$ & $40\%$ & $170$ \\
BeyondSWE CrossRepo   & $18$ & $8$  & $44\%$ & $378$ \\
BeyondSWE DepMigrate  & $25$ & $10$ & $40\%$ & $412$ \\
GAIA                  & $28$ & $15$ & $54\%$ & $75$  \\
LoCoBench Feature     & $60$ & $38$ & $63\%$ & $394$ \\
LoCoBench Refactoring & $44$ & $20$ & $45\%$ & $353$ \\
LOCA-Bench            & $68$ & $30$ & $44\%$ & $540$ \\
ResearchRubrics       & $44$ & $25$ & $57\%$ & $221$ \\
\midrule
\textbf{Total}        & $\mathbf{348}$ & $\mathbf{172}$ & $\mathbf{49\%}$ & $\mathbf{2{,}790}$ \\
\bottomrule
\end{tabular}
\end{table}

\paragraph{Cost comparison with the evolution baselines.}
We compare Meta-Team's evolution cost with two representative evolution
baselines, ReCreate and MASFly, under the same setting as the main results.
As shown in Table~\ref{tab:evolution_cost_compare}, Meta-Team's total evolution
cost (\$348) is on the same order as the baselines (\$216--\$268), about 30\%
above MASFly and 60\% above ReCreate. This difference stays bounded because
agents in Meta-Team communicate on demand, based on their own reasoning about
whether an exchange is needed, rather than exchanging all-to-all. For reference,
this level of evolution cost accompanies average gains of $+6.3$ over MASFly and
$+7.9$ over ReCreate, with statistically significant margins
(Table~\ref{tab:main_results} and Appendix~\ref{app:statistical_reliability}).

\begin{table}[h]
\centering
\small
\caption{The \emph{Evolution Run Cost} of Meta-Team versus the evolution
baselines (USD). All values are rounded to integers.}
\label{tab:evolution_cost_compare}
\begin{tabular}{@{}lrrr@{}}
\toprule
\textbf{Benchmark} & \textbf{Meta-Team} & \textbf{ReCreate} & \textbf{MASFly} \\
\midrule
SWE-Pro Ansible       & $30$ & $19$ & $21$ \\
SWE-Pro Qutebrowser   & $30$ & $17$ & $21$ \\
BeyondSWE CrossRepo   & $18$ & $14$ & $13$ \\
BeyondSWE DepMigrate  & $25$ & $19$ & $18$ \\
GAIA                  & $28$ & $18$ & $30$ \\
LoCoBench Feature     & $60$ & $33$ & $42$ \\
LoCoBench Refactoring & $44$ & $25$ & $39$ \\
LOCA-Bench            & $68$ & $47$ & $53$ \\
ResearchRubrics       & $44$ & $24$ & $31$ \\
\midrule
\textbf{Total}        & $\mathbf{348}$ & $\mathbf{216}$ & $\mathbf{268}$ \\
\bottomrule
\end{tabular}
\end{table}

\appsubsection{Generalization Across Base Models}
\label{app:additional:backbone}

\begin{figure}[h]
 \centering
    \begin{minipage}[t]{0.56\textwidth}
        \centering
      \includegraphics[width=\textwidth]{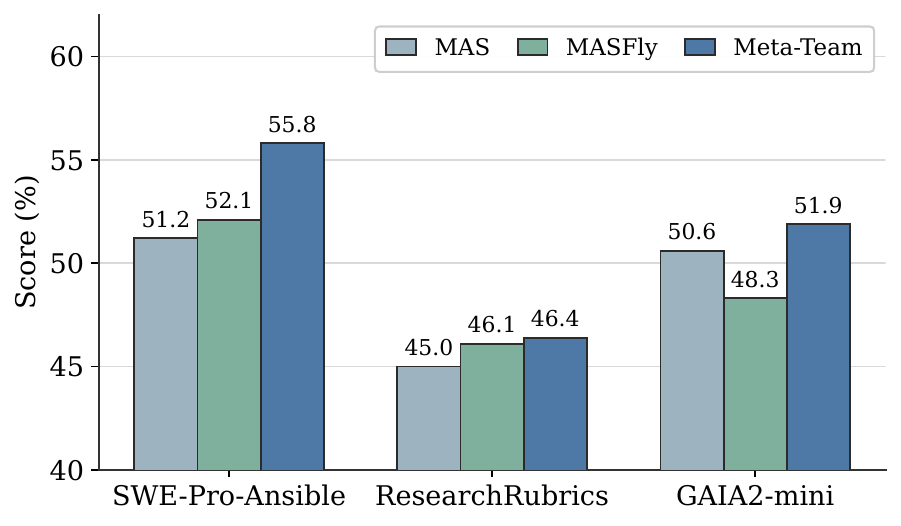}
    \captionof{figure}{Cross-model validation with GPT-5.5.}
        \label{fig:gpt55_cross_model}
    \end{minipage}
    \hfill
    \begin{minipage}[t]{0.42\textwidth}
        \centering
        \includegraphics[width=\textwidth]{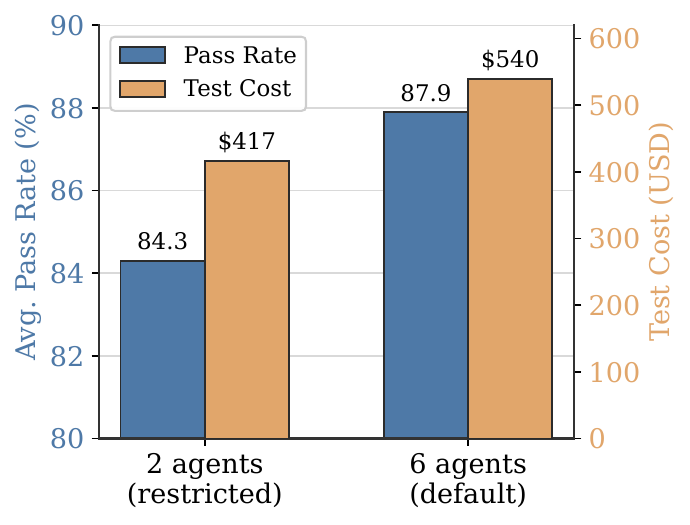}
      \captionof{figure}{Effect of team size.}
    \label{fig:team_size}
    \end{minipage}
\end{figure}

The main experiments use Claude Sonnet~4.6 as the backbone. To examine whether
the gains depend on a specific model, we re-run the full evolution and
evaluation pipeline with GPT-5.5 as the backbone on three representative
benchmarks that span our task categories: SWE-Pro-Ansible (coding, long-horizon),
ResearchRubrics (non-coding, open-ended research), and GAIA2-mini (non-coding,
tool-agentic).\footnote{GAIA2~\citep{froger2026gaia2} is a recent tool-agentic
benchmark over a simulated tool-and-app ecosystem; we use its official mini
subset with the same evolution/evaluation protocol as the main paper.}
We compare against the initial hand-crafted MAS (the starting
point of evolution) and MASFly (the strongest MAS-evolution baseline in
Table~\ref{tab:main_results}), keeping all other settings fixed. We report pass
rate (\%) for SWE-Pro-Ansible and GAIA2-mini and the average rubric score
(0--100) for ResearchRubrics. As shown in Figure~\ref{fig:gpt55_cross_model},
Meta-Team remains the best method on all three benchmarks under GPT-5.5,
indicating that its benefits carry over across frontier model families rather
than being tied to a single model.

\appsubsection{Effect of Team Size}
\label{app:additional:teamsize}

Team size in Meta-Team is not a fixed hyperparameter but an outcome of evolution:
the open-roster mechanism (Appendix~\ref{app:pipeline}) can dynamically recruit
new members, and team-level evolution (L3) can introduce new roles from
accumulated experience. This adaptation is already visible on LOCA-Bench
(\S\ref{sec:Scalability and Generalization of Meta-Team}), where evolving only on
96K-token cases leads Meta-Team to add two extra workers, which is what keeps it
effective at the longest context (256K).

To directly examine the effect of varying the number of agents, we take the same
evolved team and, at test time, restrict its active roster to 2 agents
(1 Chairman $+$ 1 Worker), comparing against the default evolved team (6 agents);
all other settings follow Figure~\ref{fig:locabench_scale}, and both pass rate
and test cost are averaged over the six context scales (8K--256K).
As shown in Figure~\ref{fig:team_size}, allowing the team to grow improves the
average pass rate by $+3.6$ while the test cost rises only moderately
($\sim$30\%), suggesting that Meta-Team scales with team size in both accuracy
and cost because the open-roster protocol recruits only the agents a task needs.

\appsubsection{Stability Across Independent Evolution Runs}
\label{app:additional:stability}

All results in the main paper are reported as avg@3 (three independent rollouts
per instance), and the main-result margins are backed by paired significance
tests (Appendix~\ref{app:statistical_reliability}). These characterize evaluation
variance, but not the variance of the evolution process itself, which is driven
by LLM reflection. To characterize the latter, we repeat the entire
self-evolution pipeline from scratch under three independent runs on the main
backbone (Claude Sonnet~4.6); each run re-executes the full L1/L2/L3 evolution
and produces its own evolved team, which is then evaluated. We use the same three
representative benchmarks and baselines as in
Appendix~\ref{app:additional:backbone}, reporting pass rate (\%) for
SWE-Pro-Ansible and GAIA2-mini and the average rubric score (0--100) for
ResearchRubrics.
Table~\ref{tab:stability_runs} reports the mean and standard deviation across the
three runs. The small deviations indicate that Meta-Team's gains are stable
across independent evolution runs rather than the product of a single evolution
trajectory. (MAS does not evolve, so its deviation reflects avg@3 evaluation
variance only.)

\begin{table}[h]
\centering
\small
\caption{Results across three independent evolution runs on the main backbone
(mean $\pm$ std).}
\label{tab:stability_runs}
\begin{tabular}{@{}lccc@{}}
\toprule
\textbf{Method} & \textbf{SWE-Pro-Ansible} & \textbf{ResearchRubrics} & \textbf{GAIA2-mini} \\
\midrule
MAS           & $42.7 \pm 1.8$ & $49.3 \pm 2.1$ & $43.8 \pm 1.9$ \\
MASFly        & $44.8 \pm 1.9$ & $50.1 \pm 2.2$ & $44.2 \pm 1.7$ \\
\textbf{Meta-Team} & $\mathbf{53.1 \pm 1.6}$ & $\mathbf{53.7 \pm 1.5}$ & $\mathbf{46.9 \pm 2.4}$ \\
\bottomrule
\end{tabular}
\end{table}



\end{document}